\documentclass[10pt,journal]{IEEEtran}

\usepackage{cite,calc} %,graphicx,url,stfloats
\usepackage{amsfonts,amsmath,amssymb,graphics,psfrag,subfigure,epsfig}
\usepackage[mathcal]{euscript}  % redefine \mathcal to be \EuScript;
\usepackage{verbatim} % get comment environment, + new verbatim
 \usepackage{amsxtra}  % get, eg, \accentedsymbol
\DeclareMathOperator*{\argmax}{arg\,max}

\DeclareMathOperator{\tr}{tr}

\DeclareMathOperator{\var}{var}
\DeclareMathOperator{\cov}{cov}

\DeclareMathOperator{\Null}{Null}

\newcounter{actr}
{\begin{list}{(\alph{actr})}{\usecounter{actr}}}{\end{list}}

\newcounter{ictr}
{\begin{list}{(\roman{ictr})}{\usecounter{ictr}}}{\end{list}}

\newtheorem{thm}{Theorem}
\newtheorem{lemma}{Lemma}
\newtheorem{claim}{Claim}
\newtheorem{corol}{Corollary}
\newtheorem{prop}{Proposition}
\newtheorem{defn}{Definition}
\newtheorem{fact}{Fact}
\newenvironment{new-proof}[1]
{{\em Proof  #1: }}%
{ \noindent\qed }

\newcommand{\cf}{cf.}

\newcommand{\qed}{\rule[0.1ex]{1.4ex}{1.6ex}}

\newcommand{\defeq}{\stackrel{\Delta}{=}}

\newcommand{\compls}{\mathbb{C}}

\newcommand{\mrm}{\mathrm}

\newcommand{\T}{{\mathrm{T}}}
\DeclareMathOperator{\R}{Re}

\DeclareMathAlphabet{\mathbsf}{OT1}{cmss}{bx}{n}% bold sans serif
\DeclareMathAlphabet{\mathssf}{OT1}{cmss}{m}{sl}% slanted sans serif

\DeclareSymbolFont{bsfletters}{OT1}{cmss}{bx}{n}  
\DeclareSymbolFont{ssfletters}{OT1}{cmss}{m}{n}
\DeclareMathSymbol{\bsfGamma}{0}{bsfletters}{'000}
\DeclareMathSymbol{\ssfGamma}{0}{ssfletters}{'000}
\DeclareMathSymbol{\bsfDelta}{0}{bsfletters}{'001}
\DeclareMathSymbol{\ssfDelta}{0}{ssfletters}{'001}
\DeclareMathSymbol{\bsfTheta}{0}{bsfletters}{'002}
\DeclareMathSymbol{\ssfTheta}{0}{ssfletters}{'002}
\DeclareMathSymbol{\bsfLambda}{0}{bsfletters}{'003}
\DeclareMathSymbol{\ssfLambda}{0}{ssfletters}{'003}
\DeclareMathSymbol{\bsfXi}{0}{bsfletters}{'004}
\DeclareMathSymbol{\ssfXi}{0}{ssfletters}{'004}
\DeclareMathSymbol{\bsfPi}{0}{bsfletters}{'005}
\DeclareMathSymbol{\ssfPi}{0}{ssfletters}{'005}
\DeclareMathSymbol{\bsfSigma}{0}{bsfletters}{'006}
\DeclareMathSymbol{\ssfSigma}{0}{ssfletters}{'006}
\DeclareMathSymbol{\bsfUpsilon}{0}{bsfletters}{'007}
\DeclareMathSymbol{\ssfUpsilon}{0}{ssfletters}{'007}
\DeclareMathSymbol{\bsfPhi}{0}{bsfletters}{'010}
\DeclareMathSymbol{\ssfPhi}{0}{ssfletters}{'010}
\DeclareMathSymbol{\bsfPsi}{0}{bsfletters}{'011}
\DeclareMathSymbol{\ssfPsi}{0}{ssfletters}{'011}
\DeclareMathSymbol{\bsfOmega}{0}{bsfletters}{'012}
\DeclareMathSymbol{\ssfOmega}{0}{ssfletters}{'012}

\renewcommand{\defeq}{\triangleq}

\newcommand{\rvbB}{{\mathbsf{B}}}	% b

\newcommand{\rve}{{\mathssf{e}}}	% e

\newcommand{\rvbh}{{\mathbsf{h}}}
\newcommand{\rvbH}{{\mathbsf{H}}}

\newcommand{\rvq}{{\mathssf{q}}}	% q

\newcommand{\rvu}{{\mathssf{u}}}	% u

\newcommand{\rvv}{{\mathssf{v}}}	% v

\newcommand{\rvbv}{{\mathbsf{v}}}

\newcommand{\rvw}{{\mathssf{w}}}	% w

\newcommand{\rvx}{{\mathssf{x}}}	% x, random variable

\newcommand{\rvbx}{{\mathbsf{x}}}

\newcommand{\rvy}{{\mathssf{y}}}	% y

\newcommand{\rvby}{{\mathbsf{y}}}

\newcommand{\rvz}{{\mathssf{z}}}	% z

\newcommand{\rvbz}{{\mathbsf{z}}}

\newcommand{\ba}{{\mathbf{a}}}

\newcommand{\bA}{{\mathbf{A}}}

\newcommand{\bB}{{\mathbf{B}}}

\newcommand{\cE}{{\mathcal{E}}}

\newcommand{\bg}{{\mathbf{g}}}

\newcommand{\bG}{{\mathbf{G}}}

\newcommand{\bh}{{\mathbf{h}}}

\newcommand{\bH}{{\mathbf{H}}}

\newcommand{\bI}{{\mathbf{I}}}

\newcommand{\bK}{{\mathbf{K}}}

\newcommand{\cK}{{\mathcal{K}}}

\newcommand{\CN}{{\mathcal{CN}}}
\newcommand{\cO}{{\mathcal{O}}}

\newcommand{\cP}{{\mathcal{P}}}

\newcommand{\bQ}{{\mathbf{Q}}}

\newcommand{\cS}{{\mathcal{S}}}

\newcommand{\bv}{{\mathbf{v}}}

\newcommand{\cW}{{\mathcal{W}}}

\newcommand{\bx}{{\mathbf{x}}}

\newcommand{\bz}{{\mathbf{z}}}

\newcommand{\rz}{\mathsf{{z}}}

\newcommand{\al}{\alpha}
\newcommand{\bal}{{\boldsymbol{\al}}}

\newcommand{\g}{\gamma}

\newcommand{\eps}{\varepsilon}

\newcommand{\bph}{{\boldsymbol{\phi}}}

\newcommand{\bps}{{\boldsymbol{\psi}}}

\newcommand{\bth}{{\boldsymbol{\theta}}}

\newcommand{\La}{\Lambda}
\newcommand{\bLa}{{\boldsymbol{\La}}}

\newcommand{\mpf}{\xi}
\newcommand{\Nt}{n_\mrm{t}}
\newcommand{\Ne}{n_\mrm{e}}
\newcommand{\lamax}{\lambda_\mrm{max}}
\newcommand{\bpsmax}{\bps_\mrm{max}}
\newcommand{\bhr}{\bh_\mrm{r}}
\newcommand{\tbhr}{\tilde{\bh}_\mrm{r}}
\newcommand{\bgr}{\bg_\mrm{r}}
\newcommand{\rvbhr}{\rvbh_\mrm{r}}
\newcommand{\trvbhr}{\tilde{\rvbh}_\mrm{r}}
\newcommand{\bHe}{\bH_\mrm{e}}

\newcommand{\bGe}{\bG_\mrm{e}}
\newcommand{\rvbHe}{\rvbH_\mrm{e}}
\newcommand{\trvbHe}{\tilde{\rvbH}_\mrm{e}}

\begin{document}
%
% paper title

% \title{Secure Space-Time Coding}
\title{Secure Transmission with Multiple Antennas:\\ 
       The MISOME Wiretap Channel}

\author{Ashish~Khisti,~\IEEEmembership{Student~Member,~IEEE,}
        and~Gregory~W.~Wornell,~\IEEEmembership{Fellow,~IEEE}% <-this % stops a space
\thanks{Manuscript received August 2007.
This work was supported in part by NSF under Grant No.~CCF-0515109.
        This work was presented in part at the International Symposium
        on Information Theory (ISIT), Nice, France, June 2007.}%
\thanks{The authors are with the Department of Electrical Engineering
        and Computer Science, Massachusetts Institute of Technology,
        Cambridge, MA~~02139.  (Email: \{khisti,gww\}@mit.edu.)}}

\maketitle

\begin{abstract}
The role of multiple antennas for secure communication is investigated
within the framework of Wyner's wiretap channel.  We characterize the
secrecy capacity in terms of generalized eigenvalues when the sender
and eavesdropper have multiple antennas, the intended receiver has a
single antenna, and the channel matrices are fixed and known to all
the terminals, and show that a beamforming strategy is
capacity-achieving.  In addition, we show that in the high
signal-to-noise (SNR) ratio regime the penalty for not knowing
eavesdropper's channel is small---a simple ``secure space-time code''
that can be thought of as masked beamforming and radiates power
isotropically attains near-optimal performance.  In the limit of large
number of antennas, we obtain a realization-independent
characterization of the secrecy capacity as a function of the number
$\beta$: the number of eavesdropper antennas per sender antenna.  We show that the
eavesdropper is comparatively ineffective when $\beta<1$, but that for
$\beta\ge2$ the eavesdropper can drive the secrecy capacity to zero,
thereby blocking secure communication to the intended receiver.
Extensions to ergodic fading channels are also provided.
\end{abstract}


\IEEEpeerreviewmaketitle

\section{Introduction}
\label{sec:intro}

\PARstart{M}{ultiple-element} antenna arrays are finding growing use in
wireless communication networks.  Much research to date has focused on
the role of such arrays in enhancing the throughput and robustness for
wireless communication systems.  By contrast, this paper focuses on
the role of such arrays in a less explored aspect of wireless
systems---enhancing security.  Specifically, we develop and optimize
physical layer techniques for using multiple antennnas to protect
digital transmissions from potential eavesdroppers, and analyze the
resulting performance characteristics.

A natural framework for protecting information at the physical layer
is the so-called wiretap channel introduced by
Wyner~\cite{wyner:75Wiretap} and associated notion of secrecy
capacity.  In the basic wiretap channel, there are three
terminals---one sender, one receiver and one eavesdropper.  Wyner's
original treatment established the secrecy capacity for the case where
the underlying broadcast channel between the sender and the receiver
and eavesdropper is a degraded one.  Subsequent work generalized this
result to nondegraded discrete memoryless broadcast
channels~\cite{csiszarKorner:78}, and applied it to the basic Gaussian
channel~\cite{leung-Yan-CheongHellman:99}.

Motivated by emerging wireless communication applications, there is
growing interest in extending the basic Gaussian wiretap channel to
the case when the terminals have multiple antennas; see, e.g.,
\cite{NegiGoel05, GoelNegi05, ParadaBlahut:05, khistiTchamWornell:06,
gopalaLaiElGamal:06Secrecy, khistiTchamWornell:07,
khistiWornellEldar:07, LiTrappeYates07, shaifeeUlukus:07} and the
references therein.  While in principle the secrecy capacity for such
nondegraded broadcast channels is developed in~\cite{csiszarKorner:78}
by Csisz{\'a}r and K{\"o}rner, the solution is in terms of an
optimized auxiliary random variable and has been prohibitively
difficult to explicitly evaluate.  Thus, such characterizations of the
solution have not proved particularly useful in practice.

In this paper, we investigate practical characterizations for the
specific scenario in which the sender and eavesdropper have multiple
antennas, but the intended receiver has a single antenna.  We refer to
this configuration as the multi-input, single-output,
multi-eavesdropper (MISOME) case.  It is worth emphasizing that the
multiple eavesdropper antennas can correspond to a physical
multiple-element antenna array at a single eavesdropper, a collection
of geographically dispersed but perfectly colluding single-antenna
eavedroppers, or related variations.

We first develop the secrecy capacity when the complex channel gains
are fixed and known to all the terminals.  A novel aspect of our
derivation is our approach to (tightly) upper bounding the secrecy
capacity for the wiretap channel. Our result thus
indirectly establishes the optimum choice of auxiliary random variable
in the secrecy capacity expression of ~\cite{csiszarKorner:78},
addressing an open problem.

While the capacity achieving scheme generally requires that the the
sender and the intended receiver have knowledge of the eavesdropper's
channel (and thus number of antennas as well)---which is often not
practical---we futher show that performance is not strongly sensitive
to this knowledge.  Specifically, we show that a simple \emph{masked
beamforming} scheme described in~\cite{NegiGoel05,GoelNegi05} that
does not require knowledge of the eavesdropper's channel is close to
optimal in the high SNR regime.

In addition, we examine the degree to which the eavesdropper can drive
the secrecy capacity of the channel to zero, thereby effectively
blocking secure communication between sender and (intended)
receiver.  In particular, for Rayleigh fading in the large antenna
array limit, we use random matrix theory to characterize the secrecy
capacity (and the rate achievable by masked beamforming) as a function
of the ratio of the number of antennas at the eavesdropper to that at
the sender.  Among other results in this scenario, we show that 1) to
defeat the security in the transmission it is sufficient for the
eavesdropper to use at least twice as many antennas as the sender; and
2) an eavesdropper with significantly fewer antennas than the
transmitter is not particularly effective.

Our results extend to the case of time-varying channels.  We focus on
the case of fast (ergodic, Rayleigh) fading, where the message is
transmitted over a block that is long compared to the coherence time
of the fading.  In our model the state of the channel to the receiver
is known by all three parties (sender, receiver, and eavesdropper),
but the state of the channel to the eavesdropper is known only to the
eavesdropper.  Building on techniques developed for the single
transmitter antenna wiretap
problems~\cite{khistiTchamWornell:07,gopalaLaiElGamal:06Secrecy}, we
develop upper and lower bounds on the secrecy capacity both for
finitely many antennas and in the large antenna limit.

As a final comment, we note that the idea of protecting information at
the physical layer (rather than the application layer) is not a
conventional approach in contemporary cryptography.  Indeed, the
common architecture today has the lower network layers focus on
providing a noiseless public bit-pipe and the higher network layers
focus on enabling privacy via the exchange and distribution of
encryption keys among legitimate parties prior to the commencement of
communication.  As discussed in
\cite{khistiTchamWornell:07,khistiTchamWornell:06}, for many emerging
applications, existing key distribution methods are difficult to
exploit effectively.  In such cases, physical-layer mechanisms such as
those developed in this paper constitute a potentially attractive
alternative approach to providing transmission security.

The organization of the paper is as follows.
Section~\ref{sec:notation} summarizes some convenient notation 
used in the paper and some mathematical preliminaries.  Section~\ref{sec:ChModel}
describes the channel and system model of interest.
Section~\ref{sec:results} states all the main results of the paper.
The proofs of our results appear in subsequent sections and the more
technical details are provided in the Appendices.
Section~\ref{sec:UpperBound} provides an alternate upper bound while
Section~\ref{sec:MISO} provides the secrecy capacity.  Our analysis of
the masked beamforming scheme is provided in Section~\ref{sec:AN}
while the scaling laws of the secrecy capacity and the masked
beamforming scheme are provided in section~\ref{sec:scaling}.  The
extension to ergodic fading channels with only intended receiver's
channel state information is treated in
Section~\ref{sec:FadingChannels} and Section~\ref{sec:Conclusion}
contains some concluding remarks.

\section{Notation}
\label{sec:notation}

Bold upper and lower case characters are used for matrices and
vectors, respectively.  Random variables are distinguished from
realizations by the use of san-serif fonts for the former and seriffed
fonts for the latter.  And we generally reserve the symbols $I$ for
mutual information, $H$ for entropy, and $h$ for differential entropy.
All logarithms are base-2 unless otherwise indicated.

The set of all $n$-dimensional complex-valued vectors is denoted by
$\compls^n$, and the set of $m\times n$-dimensional matrices is
denoted using $\compls^{m\times n}$.  Matrix transposition is denoted
using the superscript ${\ }^\T$, and the Hermitian (i.e., conjugate)
transpose of a matrix is denoted using the superscript ${\ }^\dagger$.
Moreover, $\Null(\cdot)$ denotes the null space of its matrix
argument, and $\tr(\cdot)$ and $\det(\cdot)$ denote the trace and
determinant of a matrix, respectively.  The notation $\bA\succeq0$
means that $\bA$ is a positive semidefinite matrix and we reserve the symbol $\bI$ to denote the identity matrix, whose dimensions will  be clear from the context.

A sequence of length $n$ is either denoted by $\{x(t)\}_{t=1}^n$ or
sometimes more succinctly as $x^n$; in addition, we sometimes need
notation the $x_i^j$ for a sequence $x_i,x_{i+1},\dots,x_j$.

Finally, $\CN(0,\bK)$ denotes a zero-mean circularly-symmetric complex
Gaussian distribution with covariance $\bK$, and we use the notation
$\{\cdot\}^+ \defeq \max(0,\cdot)$ throughout the paper.

\subsection{Preliminaries: Generalized Eigenvalues}
\label{sec:prelim}

Many of our results arise out of generalized eigenvalue analysis.  We summarize the properties of generalized eigenvalues
and eigenvectors we require in the sequel.  For more extensive
developments of the topic, see, e.g., \cite{golubVanLoan,lapack}.

\begin{defn}[Generalized eigenvalues]
\label{def:Geig}
For a Hermitian matrix $\bA \in \compls^{n\times n}$ and positive
definite\footnote{When $\bB$ is singular, we replace $\lambda$ with a
pair $(\alpha,\beta)$ that satisfies $\beta \bA\bps = \alpha \bB\bps$.
A solution for which $\alpha \neq 0$ and $\beta=0$ corresponds to an
infinite eigenvector.  Generalized eigenvalues and eigenvectors also
arise in simultaneous diagonalization of
$(\bA,\bB)$~\cite{golubVanLoan}.  
\iffalse
If $\bps_1,\bps_2,\ldots,
\bps_n$ are the $n$ generalized eigenvectors of $(A,B)$, let $K =
[\bps_1, \bps_2,\ldots, \bps_n] \in \compls^{n\times n}$ then $K^H A K
= \bLa_A$, and $K^H B K = \bLa_B$, where $\bLa_A =
\mrm{diag}\{\alpha_1,\alpha_2,\ldots, \alpha_n\}$ and $\bLa_B =
\mrm{diag}\{\beta_1,\beta_2,\ldots, \beta_n\}$.  The generalized
eigenvalues are $\lambda_j = \frac{\alpha_j}{\beta_j}$, for
$j=1,2,\ldots, n$.  }
\fi
} matrix $\bB \in \compls^{n\times n}$, we refer to $(\lambda,\bps)$
as a generalized eigenvalue-eigenvector pair of $(\bA,\bB)$ if
$(\lambda,\bps)$ satisfy
\begin{equation}
\bA\bps = \lambda \bB \bps.
\label{eq:GenEig} 
\end{equation}
\end{defn}
~

Since $\bB$ in Definition~\ref{def:Geig} is invertible, first note
that generalized eigenvalues and eigenvectors can be readily expressed
in terms of regular ones.  Specifically,
\begin{fact}
The generalized eigenvalues and eigenvectors of the pair $(\bA,\bB)$
are the regular eigenvalues and eigenvectors of the matrix
$\bB^{-1} \bA$.  
\end{fact}

Other characterizations reveal more useful properties for our
development.  For example, we have the following:
\begin{fact}[Variational Characterization]
\label{fact:variationalGEig}
The generalized eigenvectors of $(\bA,\bB)$ are the stationary point
solution to a particular Rayleigh quotient.  Specifically, the largest
generalized eigenvalue is the maximum of the Rayleigh
quotient\footnote{Throughout the paper we use $\lamax$ to
denote the largest eigenvalue.  Whether this is a regular or
generalized eigenvalue will be clear from context, and when there is a
need to be explicit, the relevant matrix or matrices will be indicated
as arguments.}
\begin{equation}
\lamax(\bA,\bB) = 
\max_{\bps\in \compls^n}\frac{\bps^\dagger \bA\bps}{\bps^\dagger \bB \bps},
\label{eq:GenEigRayleigh}
\end{equation}
and the optimum is attained by the eigenvector corresponding to
$\lamax(\bA,\bB)$.
\end{fact}

The case when $\bA$ has rank one is of special interest to us.  In
this case, the generalized eigenvalue admits a particularly simple
expression:
\begin{fact}[Quadratic Form]
\label{fact:Qform} 
When $\bA$ in Definition~\ref{def:Geig} has rank one, i.e., $\bA = \ba
\ba^\dagger$ for some $\ba \in \compls^n$, then
\begin{equation}
\lamax(\ba\ba^\dagger,\bB)= \ba^\dagger \bB^{-1}\ba.
\label{eq:lamax-rankone}
\end{equation}
\end{fact}

\section{Channel and System Model}
\label{sec:ChModel}

The MISOME channel and system model is as follows.  We use
$\Nt$ and $\Ne$ to denote the number of sender and
eavesdropper antennas, respectively; the (intended) receiver has a
single antenna.   The signals observed at the receiver and
eavesdropper, respectively, are, for $t=1,2,\ldots$,
\begin{equation} 
\begin{aligned}
\rvy_\mrm{r}(t) &= \bhr^\dagger \rvbx(t) + \rvz_\mrm{r}(t) \\
\rvby_\mrm{e}(t) &= \bHe \rvbx(t) + \rvbz_\mrm{e}(t),
\end{aligned}
\label{eq:ChModel}
\end{equation}
where $\rvbx(t)\in\compls^{\Nt}$ is the transmitted signal vector,
$\bhr\in \compls^{\Nt}$ and $\bHe \in \compls^{\Ne
\times \Nt}$ are complex channel gains, and $\rvz_\mrm{r}(t)$ and
$\rvbz_\mrm{e}(t)$ are independent identically-distributed (i.i.d.)
circularly-symmetric complex-valued Gaussian noises:
$\rvz_\mrm{r}(t) \sim \CN(0,1)$ and $\rvbz_\mrm{e}(t) \sim
\CN(0,\bI)$.  Moreover, the noises are independent, and the input
satisfies an average power constraint of $P$, i.e.,
\begin{equation} 
E\left[\frac{1}{n}\sum_{t=1}^n \|\rvbx(t)\|^2\right]\le P.
\end{equation}
Finally, except when otherwise indicated, all channel gains are fixed
throughout the entire transmission period, and are known to all the
terminals.

Communication takes place at a rate $R$ in bits per channel use over a
transmission interval of length $n$.  Specifically, a $(2^{nR},n)$
code for the channel consists of a message $\rvw$ uniformly
distributed over the index set $\cW_n = \{1,2,\ldots, 2^{nR}\}$, an
encoder $\mu_n: \cW_n \rightarrow \compls^{\Nt \times n}$ that maps
the message $\rvw$ to the transmitted (vector) sequence
$\{\rvbx(t)\}_{t=1}^n$, and a decoding function $\nu_n: \compls^{n}
\rightarrow \cW_n$ that maps the received sequence
$\{y_\mrm{r}(t)\}_{t=1}^n$ to a message estimate $\hat{\rvw}$.  The
error event is $\cE_n = \{\nu_n(\mu_n(\rvw)) \neq \rvw\}$, and the
amount of information obtained by the eavesdropper from the
transmission is measured via the equivocation
$I(\rvw;\rvby_\mrm{e}^n)$.

\begin{defn}[Secrecy Capacity] 
\label{def:seccap}
A secrecy rate $R$ is achievable if there exists a sequence of
$(2^{nR},n)$ codes such that $\Pr(\cE_n)\rightarrow 0$ and
$I(\rvw;\rvby_\mrm{e}^n)/n\rightarrow 0$ as $n\rightarrow\infty$.
The \emph{secrecy capacity} is the supremum of all achievable
secrecy-rates.
\end{defn}

Note that our notion of secrecy capacity follows
\cite{wyner:75Wiretap,csiszarKorner:78,leung-Yan-CheongHellman:99} in
requiring a vanishing \emph{per-symbol} mutual information for the
eavesdropper's channel (hence the normalization by $n$ in
Definition~\ref{def:seccap}). Practically, this means that while the
eavesdropper is unable to decode any fixed fraction of the message
bits, it does not preclude the possibility of decoding a fixed
number (but vanishing fraction) of the message bits.

Maurer and Wolf~\cite{MaurerWolf00} (see also~\cite{csiszar:96}) have
observed that for discrete memoryless channels, the secrecy capacity
is not reduced even when one imposes the stronger requirement that
$I(\rvw;\rvby_\mrm{e}^n)\rightarrow 0$ as $n\rightarrow\infty$.
However, we remark in advance that it remains an open question whether
a similar result holds for the Gaussian case of interest in this work.

\section{Main Results}
\label{sec:results}

The MISOME wiretap channel is a nondegraded broadcast channel.  In
Csisz{\'a}r and K{\"o}rner \cite{csiszarKorner:78}, the secrecy
capacity of the nondegraded discrete memoryless broadcast channel
$p_{\rvy_\mrm{r},\rvby_\mrm{e}|\rvx}$ is expressed in the form
\begin{align} 
C = \max_{p_{\rvu},p_{\rvx|\rvu}}
I(\rvu;\rvy_\mrm{r})-I(\rvu;\rvy_\mrm{e}),
\label{eq:CK}
\end{align}
where $\rvu$ is an auxiliary random variable over a certain alphabet
that satisfies the Markov relation $\rvu\leftrightarrow \rvx
\leftrightarrow (\rvy_\mrm{r},\rvy_\mrm{e})$.  Moreover, the secrecy
capacity \eqref{eq:CK} readily extends to the continuous alphabet case
with a power constraint, so it also gives a characterization of the
MISOME channel capacity.

Rather than attempting to solve for the optimal choice of $\rvu$ and
$p_{\rvx|\rvu}$ in \eqref{eq:CK} directly to evaluate this
capacity,\footnote{The direct approach is explored in, e.g.,
\cite{LiTrappeYates07} and \cite{shaifeeUlukus:07}, where the
difficulty of performing this optimization is reported even when
restricting $p_{\rvx|\rvu}$ to be singular (a deterministic mapping)
and/or the input distribution to be Gaussian.} we consider an indirect
approach based on a useful upper bound as the converse, which we
describe next.  We note in advance that, as described in
\cite{khistiWornellEldar:07}, our upper bound has the added benefit
that it extends easily to the MIMOME case (i.e., when the receiver has
multiple antennas).

\subsection{Upper Bound on Achievable Rates}
\label{sec:ub}

A key result is the following upper bound, which we derive in
Section~\ref{sec:UpperBound}. 
\begin{thm}
\label{thm:ub} 
An upper bound on the secrecy capacity for the MISOME channel model is
\begin{equation}
R_+=\min_{\bK_\bph\in \cK_\bph}\max_{\bK_P\in\cK_P}R_+(\bK_P,\bK_\bph),
\label{eq:ub}
\end{equation}
where $R_+(\bK_P,\bK_\bph) = I(\rvbx;\rvy_\mrm{r}|\rvby_\mrm{e})$
with $\rvbx \sim\CN(0,\bK_P)$ and
\begin{equation} 
\cK_P \defeq \left\{\bK_P \Biggm| \bK_P \succeq 0,\quad
\tr(\bK_P) \le P \right\},
\label{eq:KP-def}
\end{equation}
and where
\begin{equation} 
\begin{bmatrix} 
\rvz_\mrm{r} \\ \rvbz_\mrm{e}
\end{bmatrix} \sim\CN(0,\bK_\bph)
\end{equation}
with
\begin{equation} 
\begin {aligned}
\cK_\bph &\defeq \left\{\bK_\bph \Biggm| \bK_\bph =
\begin{bmatrix} 1 & \bph^\dagger \\ \bph & \bI \end{bmatrix},\quad \bK_\bph \succeq 0
\right\} \\
&= \left\{\bK_\bph \Biggm| \bK_\bph =
\begin{bmatrix} 1 & \bph^\dagger \\ \bph & \bI \end{bmatrix},\quad
\|\bph\|\le1\right\}. 
\end {aligned}
\label{eq:Kph-def}
\end{equation}
\end{thm}
~

To obtain this bound, we consider a genie-aided channel in which the
eavesdropper observes $\rvby_\mrm{e}$ but the receiver observes
\emph{both} $\rvy_\mrm{r}$ and $\rvby_\mrm{e}$ .  Such a channel
clearly has a capacity larger than the original channel.  Moreover,
since it is a degraded broadcast channel, the secrecy capacity of the
genie-aided channel can be easily derived and is given by
(\cf~\cite{wyner:75Wiretap}) $\max
I(\rvbx;\rvy_\mrm{r}|\rvby_\mrm{e})$ where the maximum is over the
choice of input distributions.  As we will see, it is straightforward to establish that the maximizing input
distribution is Gaussian (in contrast to the original channel).

Next, while the secrecy capacity of the original channel depends only
on the marginal distributions $p_{\rvy_\mrm{r}|\rvbx}$ and
$p_{\rvby_\mrm{e}|\rvbx}$ (see, e.g.,~\cite{csiszarKorner:78}), mutual
information $I(\rvbx;\rvy_\mrm{r}|\rvby_\mrm{e})$ for the genie-aided
channel depends on the joint distribution
$p_{\rvy_\mrm{r},\rvby_\mrm{e}|\rvbx}$.  Accordingly we obtain the
tightest such upper bound by finding the joint distribution (having
the required marginal distributions), whence
\eqref{eq:ub}.

The optimization \eqref{eq:ub} can be carried out analytically,
yielding an explicit expression, as we now develop.  

\subsection{MISOME Secrecy Capacity}
\label{sec:seccap}

The upper bound described in the preceding section is achievable,
yielding the MISOME channel capacity.  Specifically, we have the
following theorem, which we prove in Section~\ref{sec:MISOMEcap-proof}.
\begin{thm}
\label{thm:MISOMEcap} 
The secrecy capacity of the channel \eqref{eq:ChModel} is
\begin{equation}
C(P) = \left\{ \log\lamax \left( 
\bI + P \bhr\bhr^\dagger, 
\bI + P \bHe^\dagger \bHe \right) \right\}^+,
\label{eq:MISOMEcap}
\end{equation}
with $\lamax$ denoting the largest generalized eigenvalue of its
argument pair.  Furthermore, the capacity is obtained by beamforming
(i.e., signaling with rank one covariance) along the direction
$\bpsmax$ of the\footnote{If there is more than one generalized
eigenvector for $\lamax$, we choose any one of them.}  generalized
eigenvector corresponding to $\lamax$ with an encoding of the message
using a code for the scalar Gaussian wiretap channel.
\end{thm}

We emphasize that the beamforming direction in
Theorem~\ref{thm:MISOMEcap} for achieving capacity will in general
depend on all of the target receiver's channel $\bhr$, the
eavesdropper's channel $\bHe$, and the SNR ($P$).

In the high SNR regime, the MISOME capacity \eqref{eq:MISOMEcap}
exhibits one of two possible behaviors, corresponding to whether
\begin{equation} 
\lim_{P\rightarrow\infty} C(P) = 
\left\{ \log\lamax \left( 
\bhr\bhr^\dagger, 
\bHe^\dagger \bHe \right) \right\}^+,
\label{eq:cap-high-SNR}
\end{equation}
is finite or infinite, which depends on whether or not $\bhr$
has a component in the null space of $\bHe$.  Specifically, we
have the following corollary, which we prove in
Section~\ref{sec:high-SNR-proof}. 
\begin{corol}
\label{corol:high-SNR} 
The high SNR asymptote of the secrecy capacity \eqref{eq:MISOMEcap} takes
the form
\begin{subequations}
\label{eq:high-SNR}
\begin{equation} 
\lim_{P\rightarrow\infty} 
C(P) = \{\log \lamax
   (\bhr\bhr^\dagger,\bHe^\dagger\bHe)
       \}^+\!<\!\infty
\text{\ \ if $\bHe^\perp \bhr\!=\!\mathbf{0}$},
\label{eq:high-SNR-a}
\end{equation}
\begin{equation} 
\lim_{P\rightarrow\infty} 
\left[C(P)-\log P\right] = \log\|\bHe^\perp\bhr\|^2
\text{\ \ \ if $\bHe^\perp \bhr\ne\mathbf{0}$},
\label{eq:high-SNR-b}
\end{equation}
\end{subequations}
where $\bHe^\perp$ denotes the projection matrix onto the null space
of $\bHe$.\footnote{That is, the columns of $\bHe^\perp$ constitute an
orthogonal basis for the null space of $\bHe$.}
\end{corol}

This behavior can be understood rather intuitively.  In particular,
when $\bHe^\perp \bhr = \mathbf{0}$, as is typically
the case when the eavesdropper uses enough antennas ($\Ne\ge\Nt$) or
the intended receiver has an otherwise unfortunate channel, the
secrecy capacity is SNR-limited.  In essence, while more transmit
power is advantageous to communication to the intended receiver, it is
also advantageous to the eavesdropper, resulting in diminishing
returns.  

By contrast, when $\bHe^\perp \bhr \ne \mathbf{0}$,
as is typically the case when, e.g., the eavesdropper uses
insufficiently many antennas ($\Ne<\Nt$) unless the eavesdropper 
has an otherwise unfortunate channel, the transmitter is able to steer
a null to the eavesdropper without simultaneously nulling the receiver
and thus capacity grows by 1 b/s/Hz with every 3 dB increase in
transmit power as it would if there were no eavesdropper to contend
with.

The MISOME capacity \eqref{eq:MISOMEcap} is also readily specialized
to the low SNR regime, as we develop in
Section~\ref{sec:low-SNR-proof}, and takes the following form.
\begin{corol}
\label{corol:low-SNR}
The low SNR asymptote of the secrecy capacity is
\begin{equation}
\lim_{P\rightarrow 0} 
\frac{C(P)}{P} = 
\frac{1}{\ln2}\{\lamax(\bhr\bhr^\dagger-\bHe^\dagger
\bHe)\}^+.   
\label{eq:low-SNR}
\end{equation}
\end{corol}
~

In this low SNR regime, the direction of optimal beamforming vector
approaches the (regular) eigenvector corresponding to the largest
(regular) eigenvalue of $\bhr\bhr^\dagger -
\bHe^\dagger \bHe$.  Note that the optimal direction
is in general not along $\bhr$.\footnote{ The optimal
direction is $\bhr$ in some special cases, such as if
$\bhr$ happens to be an eigenvector of
$\bHe^\dagger\bHe$.  The latter happens when, e.g.,
the $\Nt$ columns of $\bHe$ are orthogonal and have the same
norm.}  Thus, ignoring the eavesdropper is in general not an optimal
strategy even at low SNR.\@

\subsection{Eavesdropper-Ignorant Coding: Masked Beamforming}
\label{sec:mb}

In our basic model the channel gains are fixed and known to all the
terminals.  Our capacity-achieving scheme in
Theorem~\ref{thm:MISOMEcap} uses the knowledge of $\bHe$ for
selecting the beamforming direction.  However, in many applications it
may be difficult to know the eavesdropper's channel.  Accordingly, in
this section we analyze a simple alternative scheme that uses only
knowledge of $\bhr$ in choosing the transmit directions, yet
achieves near-optimal performance in the high SNR regime.

The scheme we analyze is a masked beamforming scheme described in
\cite{NegiGoel05,GoelNegi05}.  In this scheme, the transmitter signals
isotropically (i.e., with a covariance that is a scaled identity
matrix), and as such can be naturally viewed as a ``secure space-time
code.''  More specifically, it simultaneously transmits the message
(encoded using a scalar Gaussian wiretap code) in the direction
corresponding to the intended receiver's channel $\bhr$ while
transmitting synthesized spatio-temporal white noise in the orthogonal
subspace (i.e., all other directions).

The performance of masked beamforming is given by the following
proposition, which is proved in Section~\ref{sec:RMB-proof}.
\begin{prop}[Masked Beamforming Secrecy Rate]
A rate achievable by the masked beamforming scheme for the MISOME
channel is
\begin{align}
R&_\mrm{MB}(P) = \notag\\
&\!\left\{
\log\lamax
\left(\frac{P}{\Nt}\bhr\bhr^\dagger,
\bI\!+\!\frac{P}{\Nt}\bHe^\dagger\bHe\right)
   \!+\! \log\left(1\!+\!\frac{\Nt}{P \|\bhr\|^2}\right)\right
\}^+\!.
\label{eq:RMB}
\end{align}
\label{prop:RMB}
\end{prop}

While the rate \eqref{eq:RMB} is, in general, suboptimal, it
asymptotically near-optimal in the following sense, as developed in
Section~\ref{sec:RMBloss-proof}.
\begin{thm}
\label{thm:RMBloss}
The rate $R_\mrm{MB}(P)$ achievable by
masked beamforming scheme for the MISOME case [\cf\ \eqref{eq:RMB}] 
satisfies
\begin{equation}
\lim_{P\rightarrow\infty} \left[
C\left(\frac{P}{\Nt}\right) - R_\mrm{MB}(P) \right] = 0.
\label{eq:RMBloss}
\end{equation} 
\end{thm}
~

From the relation in \eqref{eq:RMBloss} we note that, in the high SNR
regime, the masked beamforming scheme achieves a rate of $C(P/\Nt)$,
where $\Nt$ is the number of transmit antennas.  Combining
\eqref{eq:RMBloss} with \eqref{eq:high-SNR}, we see that the
asymptotic masked beamforming loss is at most $\log\Nt$ b/s/Hz, or
equivalently $10\log_{10}\Nt$ dB in SNR.\@ Specifically,
\begin{equation} 
\lim_{P\rightarrow\infty} \left[ C(P) - R_\mrm{MB}(P) \right] = 
\begin{cases} \log\Nt, & \quad \bHe^\perp \bhr \neq
  \mathbf{0} \\
0, & \quad \bHe^\perp \bhr = \mathbf{0}.
\end{cases}
\label{eq:RMBLossExplicit}
\end{equation}

That at least some loss (if vanishing) is associated with the masked
beamforming scheme is expected, since the capacity-achieving scheme
performs beamforming to concentrate the transmission along the optimal
direction, whereas the masked beamforming scheme uses isotropic
inputs.

As one final comment, note that although the covariance structure of
the masked beamforming transmission does not depend on the
eavesdropper's channel, the rate of the base (scalar Gaussian wiretap)
code does, as \eqref{eq:RMB} reflects.  In practice, the selection of
this rate determines an insecurity zone around the sender, whereby the
transmission is secure from eavesdroppers outside this zone, but
insecure from ones inside.

\subsection{Example}
\label{sec:example}

In this section, we illustrate the preceding results for a typical
MISOME channel.  In our example, there are $\Nt=2$ transmit antennas,
and $\Ne=2$ eavesdropper antennas.  The channel to the receiver is
\begin{equation*} 
\bhr = \begin{bmatrix} 0.0991 + j0.8676 & 1.0814 - j1.1281
 \end{bmatrix}^\T,
\end{equation*}
while the channel to the eavesdropper is
\begin{equation} 
\bH_{\mrm{e},1}  = \begin{bmatrix}
0.3880 + j1.2024 &  -0.9825 + j0.5914 \\
0.4709 - j0.3073 &   0.6815 - j0.2125
		\end{bmatrix},
\label{eq:He-eg}
\end{equation}
where $j=\sqrt{-1}$.

\begin{figure}[tbp]
\centerline{\includegraphics[width=3.75in]{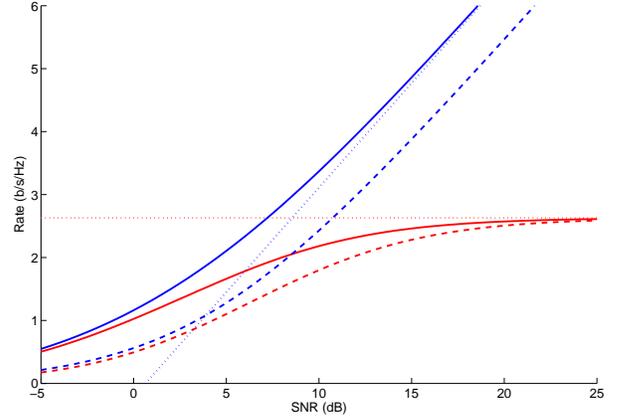}}
\caption{Performance over an example MISOME channel with $\Nt=2$
transmit antennas. The successively lower solid curves give the
secrecy capacity for $\Ne=1$ and $\Ne=2$ eavesdropper antennas,
respectively and the dotted curves indicat the corresponding high-SNR asymptote.  The dashed curves give the corresponding rates achievable
by masked beamforming, which does not require the transmitter to have
knowledge of the eavesdropper's channel. \label{fig:RMB}}
\end{figure}

Fig.~\ref{fig:RMB} depicts communication rate as a function of SNR.\@
The upper and lower solid curves depict the secrecy capacity
\eqref{eq:MISOMEcap} when the eavesdropper is using one or both its
antennas, respectively.\footnote{When a single eavesdropper antenna is
in use, the relevant channel corresponds to the first row of
\eqref{eq:He-eg}.}  As the curves reflect, when the eavesdropper has
only a single antenna, the transmitter can securely communicate at any
desired rate to its intended receiver by using enough power.  However,
by using both its antennas, the eavesdropper caps the rate at which
the transmitter can communicate securely regardless of how much power
it has available.  Note that the lower and upper curves are
representative of the cases where $\bHe^\perp \bhr$
is, and is not $\mathbf{0}$, respectively.

Fig.~\ref{fig:RMB} also shows other curves of interest.  In
particular, using dotted curves we superimpose the secrecy capacity
high-SNR asymptotes as given by \eqref{eq:high-SNR}.  As is apparent, these asymptotes
can be quite accurate approximations even for moderate values of
SNR.\@ Finally, using dashed curves we show the rate \eqref{eq:RMB}
achievable by the masked beamforming coding scheme, which doesn't use
knowledge of the eavesdropper channel.  Consistent with
\eqref{eq:RMBLossExplicit}, the loss in performance at high SNR approaches 3
dB when the eavesdropper uses only one of its antennas, and 0 dB when
it uses both.  Again, these are good estimates of the performance loss
even at moderate SNR.\@ Thus the penalty for ignorance of the
eavesdropper's channel can be quite small in practice.

\subsection{Scaling Laws in the Large System Limit}
\label{sec:sl}

Our analysis in Section~\ref{sec:seccap} of the scaling behavior of
capacity with SNR in the high SNR limit with a fixed number of
antennas in the system yielded several useful insights into secure
space-time coding systems.  In this section, we develop equally
valuable insights from a complementary scaling.  In particular, we
consider the scaling behavior of capacity with the number of antennas
in the large system limit at a fixed SNR.\@ 

One convenient feature of such analysis is that for many large
ensembles of channel gains, almost all randomly drawn realizations
produce the same capacity asymptotes.  For our analysis, we restrict
our attention to an ensemble corresponding to Rayleigh fading in which
$\rvbhr$ and $\rvbHe$ are independent, and each has
i.i.d.\ $\CN(0,1)$ entries.  The realization from the ensemble is
known to all terminals prior to communication.

In anticipation of our analysis, we make the dependency of secrecy
rates on the number of transmit and eavesdropper antennas explicit in
our notation (but leave the dependency on the realization of
$\rvbhr$ and $\rvbHe$ implicit).  Specifically, we
now use $C(P,\Nt, \Ne)$ to denote the secrecy capacity, and
$R_\mrm{MB}(P, \Nt, \Ne)$ to denote the rate of the masked beamforming
scheme.  With this notation, the scaled rates of interest are
\begin{subequations} 
\begin{equation}
\tilde{C}(\g,\beta) = 
\lim_{\Nt\rightarrow\infty} C\left(P\!=\!\g/\Nt,\Nt,\Ne\!=\!\beta \Nt\right)
\label{eq:defCAgb},
\end{equation}
and
\begin{equation}
\tilde{R}_\mrm{MB}(\g,\beta) = 
\lim_{\Nt\rightarrow\infty }R_\mrm{MB}(P\!=\!\g,\Nt,\Ne\!=\!\beta \Nt).
\label{eq:RMBg}
\end{equation}
\end{subequations}

Our choice of scalings ensures that the $\tilde{C}(\g,\beta)$ and
$\tilde{R}_\mrm{MB}(\g,\beta)$ are not degenerate.  In particular,
note that the capacity scaling \eqref{eq:defCAgb} involves an SNR
normalization.  In particular, the transmitted power $P$ is reduced as
the number of transmitter antennas $\Nt$ grows so as to keep the
\emph{received} SNR remains fixed (at specified value $\gamma$)
independent of $\Nt$.  However, the scaling \eqref{eq:RMBg} is not SNR
normalized in this way.  This is because the masked beamforming
already suffers a nominal factor of $\Nt$ SNR loss [\cf\
\eqref{eq:RMBloss}] relative to a capacity-achieving system.

In what follows, we do not attempt an exact evaluation of the secrecy
rates for our chosen scalings.  Rather we find compact lower and upper
bounds that are tight in the high SNR limit.

We begin with our lower bound, which is derived in
Section~\ref{sec:lb-scaling}.
\begin{thm}[Scaling Laws]
\label{thm:scaling} 
The asymptotic secrecy capacity satisfies
\begin{equation}
\tilde{C}(\g,\beta) \stackrel{\text{a.s.}}{\ge} \left\{\log\mpf(\g,\beta)\right\}^+,
\label{eq:MarcenkoPastur}
\end{equation}
where
\begin{align} 
\mpf(\g,&\beta)= \notag\\
&\g - \frac{1}{4} \left[ \sqrt{1\!+\!
\g\left(1\!+\!\sqrt{\beta}\right)^2}-\sqrt{1\!+\!
\g\left(1\!-\!\sqrt{\beta}\right)^2} \right]^2.
\label{eq:mpf}
\end{align}
Furthermore, the same bound holds for the corresponding asymptotic
masked beamforming rate, i.e.,
\begin{equation}
\tilde{R}_\mrm{MB}(\g,\beta) \stackrel{\text{a.s.}}{\ge}
\left\{ \log\mpf(\g,\beta)\right\}^+.
\label{eq:MarcenkoPastur-R}
\end{equation}
\end{thm}

Since the secrecy rates increase monotonically with SNR, the
infinite-SNR rates constitute a useful upper bound.   As derived in
Section~\ref{sec:high-SNR-scaling}, this bound is as follows.
\begin{thm}
\label{thm:high-SNR-scaling}
The asymptotic secrecy capacity satisfies
\begin{align} 
\tilde{C}(\g,\beta) 
&\le \lim_{\Nt\rightarrow\infty} \lim_{P\rightarrow\infty} C(P,\Nt,\beta 
\Nt) \notag\\
&\stackrel{\text{a.s.}}{=} \tilde{C}(\infty,\beta) \defeq\begin{cases} 
      0 & \beta \ge 2 \\ 
      -\log(\beta-1) & 1<\beta<2 \\
      \infty & \beta\le 1.
    \end{cases}
\label{eq:MISOMEscaling-C}
\end{align}
Furthermore, the right hand side of \eqref{eq:MISOMEscaling-C} is also
an upper bound on $\tilde{R}_\mrm{MB}(\g,\beta)$, i.e.,
\begin{align} 
\tilde{R}_\mrm{MB}(\g,\beta) &\le 
 \lim_{\Nt\rightarrow\infty} \lim_{P\rightarrow\infty}
   R_\mrm{MB}(P,\Nt,\beta \Nt) \notag\\
&\stackrel{\text{a.s.}}{=} \tilde{C}(\infty,\beta)
\label{eq:MISOMEscaling-R}
\end{align}
\end{thm}
~

Note that it is straightforward to verify that the lower bound
\eqref{eq:MarcenkoPastur} is tight at high SNR, i.e., that, for all
$\beta$,
\begin{equation} 
\left\{ \log\mpf(\infty,\beta)\right\}^+ = \tilde{C}(\infty,\beta).
\label{eq:high-SNR-tight}
\end{equation}
The same argment confirms the corresponding behavior for masked
beamforming. 

\begin{figure}[tbp]
\centerline{\includegraphics[width=3.75in]{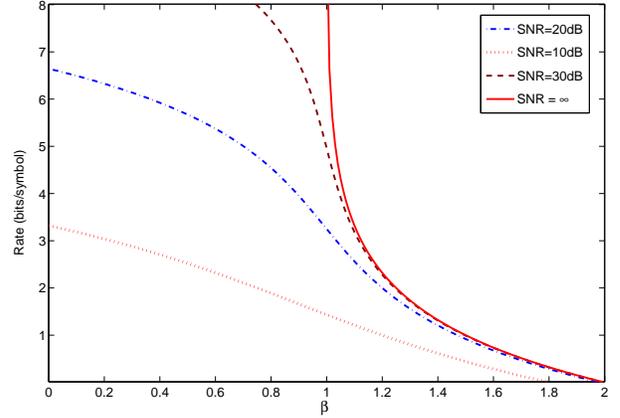}}
\caption{Secrecy capacity bounds in the large system limit.  The solid
red curve is the high SNR secrecy capacity, which is an upper bound on
the for finite SNR.\@ The progressively lower dashed curves are lower
bounds on the asymptotic secrecy capacity (and masked beamforming
secrecy rate). The channel realizations are fixed but drawn at random
according to Gaussian distribution. \label{fig:Scaling}}
\end{figure}

Our lower and upper bounds of Theorem~\ref{thm:scaling} and
Theorem~\ref{thm:high-SNR-scaling}, respectively, are depicted in
Fig.~\ref{fig:Scaling}.  In particular, we plot rate as a function of
the antenna ratio $\beta$ for various values of the SNR $\gamma$.

As Fig.~\ref{fig:Scaling} reflects, there are essentially three main
regions of behavior, the boundaries between which are increasingly
sharp with increasing SNR.\@   First, for $\beta<1$ the eavesdropper
has proportionally fewer antennas than the sender, and thus is
effectively thrwarted.   It is in this regime that the transmitter can
steer a null to the eavesdropper and achieve any desired rate to the
receiver by using enough power.

Second, for $1\le\beta<2$ the eavesdropper has proportionally more
antennas than the sender, and thus can cap the secure
rate achievable to the receiver regardless of how much power the
transmitter has available.  For instance, when the transmitter has 50\%
more antennas than the eavesdropper ($\beta=1.5$), the sender is
constrained to a maximum secure rate no more than 1 b/s/Hz.  Moreover,
if the sender is sufficiently limited in power that the received SNR
is at most, say, 10 dB, the maximum rate is less than 1/2 b/s/Hz.

We emphasize that these results imply the eavesdropper is at a
substantial disadvantage compared to the intended receiver when the
number of tranmitter antennas is chosen to be large.  Indeed, the
intended receiver needs only a single antenna to decode the message,
while the eavesdropper needs a large number of antennas to constrain
the transmission.

Finally, for $\beta\ge2$ the eavesdropper is able to entirely prevent
secure communication (drive the secrecy capacity to zero) even if the
transmitter has unlimited power available.  Useful intuition for this
phenomenon is obtained from consideration of the masked beamforming
scheme, in which the sender transmits the signal of interest in the
direction of $\bhr$ and synthesized noise in the $\Nt-1$
directions orthogonal to $\bhr$.  With such a transmission,
the intended receiver experiences a channel gain of
$\|\bhr\|^2P/\Nt$.  In the high SNR regime, the eavesdropper
must cancel the synthesized noise, which requires at least $\Nt-1$
receive antennas.  Moreover, after canceling the noise it must have
the ``beamforming gain'' of $\Nt$ so its channel quality is of the
same order as that of the intended receiver.  This requires having at
least $\Nt$ more antennas.  Thus at least $2\Nt-1$ antennas are
required by the eavesdropper to guarantee successful interception of
the transmission irrespective of the power used, which corresponds to
$\beta\ge2$ as $\Nt\rightarrow\infty$.

\subsection{Capacity Bounds in Fading}
\label{sec:fading}

Thus far we have focused on the scenarios where the receiver and
eavesdropper channels are fixed for the duration $n$ of the message
transmission.  In this section, we briefly turn our attention to the
case of time-varying channels---specifically, the case of fast fading
where there are many channel fluctuations during the course of
transmission.  In particular, we consider a model in which
$\rvbhr(t)$ and $\rvbHe(t)$ are temporally and
spatially i.i.d.\ sequences that are independent of one another
and have $\CN(0,1)$ elements, corresponding to Rayleigh fading.

In our model, $\rvbhr(t)$ is known (in a causal manner) to all
the three terminals, but only the eavesdropper has knowledge of
$\rvbHe(t)$.  Accordingly, the channel model is, for
$t=1,2,\ldots$,
\begin{equation}
\begin{aligned}
\rvy_\mrm{r}(t) &= \rvbhr^\dagger(t)\rvbx(t) + \rvz_\mrm{r}(t)\\
 \rvby_\mrm{e}(t) &= \rvbHe(t)\rvbx(t) + \rvbz_\mrm{e}(t).
\end{aligned}
\label{eq:ChModelFF}
\end{equation}

The definition of the secrecy rate and capacity is as in
Definition~\ref{def:seccap}, with the exception that the equivocation
$I(\rvw;\rvby_\mrm{e}^n)$ is replaced with
$I(\rvw;\rvby_\mrm{e}^n,\rvbHe^n|\rvbhr^n)$, which takes
into account the channel state information at the different terminals.

For this model we have the following nontrivial upper and lower bounds
on the secrecy capacity, which are developed in
Section~\ref{sec:FadingChannels}.  The upper bound is developed via
the same genie-aided channel analysis used in the proof of
Theorem~\ref{thm:MISOMEcap}, but with modifications to account for the
presence of fading.  The lower bound is achieved by the adaptive
version of masked beamforming described in \cite{NegiGoel05}.

\begin{thm}
\label{thm:RFF}
The secrecy capacity for the MISOME fast fading channel
\eqref{eq:ChModelFF} is bounded by
\begin{subequations} 
\begin{align}
C_\mrm{FF}(P,\Nt,\Ne) &\ge
\max_{\rho(\cdot)\in\cP_\mrm{FF}}E\left[ R_{\mrm{FF},-}(\rvbhr, 
\rvbHe, \rho(\cdot)) \right], \label{eq:RFF-lb} \\
C_\mrm{FF}(P,\Nt,\Ne) &\le
\max_{\rho(\cdot)\in\cP_\mrm{FF}}E\left[ R_{\mrm{FF},+}(\rvbhr, \rvbHe,
  \rho(\cdot)) \right], \label{eq:RFF-ub}
\end{align}%
\end{subequations}%
where $\cP_\mrm{FF}$ is the set of all valid power allocations, i.e.,
\begin{equation} 
\cP_\mrm{FF} = \bigl\{ \rho(\cdot) \bigm| \rho(\cdot)\ge0,\ 
E[\rho(\rvbhr)]\le P \bigr\},
\end{equation}
and
\begin{subequations} 
\begin{align}
R_{\mrm{FF},-}(\bhr, &\bHe,\rho(\cdot)) \defeq \notag\\
&\log\left(
\frac{\rho(\bhr)}{\Nt}\bhr^\dagger
\left[ 
\bI+ \frac{\rho(\bhr)}{\Nt}\bHe^\dagger \bHe
\right]^{-1}
\bhr\right) \notag\\
&\qquad + 
\log\left(1+\frac{\Nt}{\rho(\bhr)
\|\bhr\|^2}\right)\label{eq:RFF-}. \\
R_{\mrm{FF},+}(\bhr, &\bHe,\rho(\cdot)) \defeq
\notag\\
&\bigl\{\log\lamax(\bI +
\rho(\bhr)\bhr\bhr^\dagger, \bI +
\rho(\bhr)\bHe^\dagger 
\bHe)\bigr\}^+, \label{eq:RFF+}
\end{align}%
\label{eq:RFF}%
\end{subequations}%
\end{thm}

In general, our upper and lower bounds do not coincide.  Indeed, even
in the case of single antennas at all terminals ($\Nt=\Ne=1$), the
secrecy capacity for the fading channel is unknown, except in the case
of large coherence period~\cite{gopalaLaiElGamal:06Secrecy}.  

However, based on our scaling analysis in Section~\ref{sec:sl}, there
is one regime in which the capacity can be calculated: in the limit of
both high SNR and a large system.  Indeed, since
\eqref{eq:MarcenkoPastur-R} and \eqref{eq:MISOMEscaling-C} hold for
almost every channel realization, we have the following proposition,
whose proof is provided in Section~\ref{sec:scaling-FF}.
\begin{prop}
\label{prop:AsympRANFF} 
The secrecy capacity of the fast fading channel satisfies
\begin{equation}
\lim_{\Nt\rightarrow\infty}
C_\mrm{FF}(P\!=\!\g,\Nt,\Ne\!=\!\beta\Nt) {\ge} \left\{
\log\mpf(\g,\beta) \right\}^+,
\label{eq:MarcenkoPasturFF}
\end{equation}
where $\mpf(\cdot,\cdot)$ is as defined in \eqref{eq:mpf}, and
\begin{equation} 
\lim_{\Nt\rightarrow\infty}
C_\mrm{FF}(P\!=\!\g,\Nt,\Ne\!=\!\beta\Nt)  
{\le} \tilde{C}(\infty,\beta)
\label{eq:MISOMEscaling-CFF}
\end{equation}
with the $\tilde{C}(\infty,\beta)$ as given in \eqref{eq:MISOMEscaling-C}.
\end{prop}
Finally, via \eqref{eq:high-SNR-tight} we see that
\eqref{eq:MarcenkoPasturFF} and \eqref{eq:MISOMEscaling-CFF} converge
as $\g\rightarrow\infty$.

This concludes our statement of the main results.  The following
sections are devoted to the proofs of these results and some further
discussion.

\section{Upper Bound Derivation}
\label{sec:UpperBound}

In this section we prove Theorem~\ref{thm:ub}.  We begin with the
following lemma, which establishes that the capacity of genie-aided
channel is an upper bound on the channel of interest.  A proof is
provided in Appendix~\ref{app:ub}, and closely follows the general
converse of Wyner~\cite{wyner:75Wiretap}, but differs in that the
latter was for discrete channels and thus did not incorporate a power
constraint.
\begin{lemma}
\label{lem:ub}
An upper bound on the secrecy capacity of the MISOME wiretap channel
is 
\begin{equation}
C \le \max_{p_{\rvbx}\in
\cP}I(\rvbx;\rvy_\mrm{r}|\rvby_\mrm{e}),
\label{eq:Lem1UB}
\end{equation}
where $\cP$ is the set of all probability distributions that satisfy
$E[\|\rvbx\|^2]\le P$.
\end{lemma}

Among all such bounds, we can choose that corresponding to the noises
$(\rvz_\mrm{r},\rvbz_\mrm{e})$ being jointly Gaussian (they are
already constrained to be marginally Gaussian) with a covariance
making the bound as small as possible.  Then, provided the maximizing
distribution in \eqref{eq:Lem1UB} is Gaussian, we can express the
final bound in the form \eqref{eq:ub}

It thus remains only to show that the maximizing distribution is
Gaussian.
\begin{lemma}
\label{lem:GaussOpt}
For each $\bK_\bph \in \cK_\bph$, the distribution $p_{\rvbx}$ 
maximizing $I(\rvbx;\rvy_\mrm{r}|\rvby_\mrm{e})$ is Gaussian.
\end{lemma}

\begin{proof}
Since
\begin{equation*} 
I(\rvbx;\rvy_\mrm{r}|\rvby_\mrm{e}) = h(\rvy_\mrm{r} |
\rvby_\mrm{e})- h(\rvz_\mrm{r} | \rvbz_\mrm{e}),
\end{equation*}
and the second term does not depend on $p_{\rvbx}$, it suffices to
establish that $h(\rvy_\mrm{r} | \rvby_\mrm{e})$ is maximized when
$\rvbx$ is Gaussian.

To this end, let $\bal_\mrm{LMMSE}\rvby_\mrm{e}$ denote the linear
minimum mean-square error (MMSE) estimator of $\rvy_\mrm{r}$ from
$\rvby_\mrm{e}$, and $\lambda_\mrm{LMMSE}$ the corresponding mean-square
estimation error.  Recall that
\begin{align}
\bal_\mrm{LMMSE} &= (\bhr^\dagger \bK_P \bHe^\dagger
+ \bph^\dagger)(\bI + \bHe \bK_P \bHe^\dagger)^{-1},
\label{eq:ALLSE} \\ 
\lambda_\mrm{LMMSE} &= 
1 + \bhr^\dagger \bK_P \bhr \notag\\
&{}-(\bhr^\dagger \bK_P
\bHe^\dagger\!+\!\bph^\dagger)(\bI\!+\!\bHe \bK_P
\bHe^\dagger)^{-1} (\bph\!+\!\bHe \bK_P
\bhr)\label{eq:LLSE}
\end{align} 
depend on the input and noise distributions only through their (joint)
second-moment characterization, i.e., 
\begin{equation} 
\bK_P = \cov\rvbx,\qquad \bK_\bph = 
\begin{bmatrix} 1 & \bph^\dagger \\ \bph & \bI \end{bmatrix} =
\cov \begin{bmatrix} \rvz_\mrm{r}\\ \rvbz_\mrm{e} \end{bmatrix}.
\end{equation}

Proceeding, we have
\begin{align}
h(\rvy_\mrm{r} | \rvby_\mrm{e}) &= h(\rvy_\mrm{r} -
\bal_\mrm{LMMSE}\rvby_\mrm{e} | \rvby_\mrm{e}) \label{eq:mmse-1}\\ 
&\le  h(\rvy_\mrm{r} - \bal_\mrm{LMMSE}\rvby_\mrm{e}) \label{eq:mmse-2}\\
&\le \log 2\pi e \lambda_\mrm{LMMSE}, \label{eq:mmse-3}
\end{align}
where \eqref{eq:mmse-1} holds because adding a constant doesn't change
entropy, \eqref{eq:mmse-2} holds because conditioning only reduces
differential entropy, and \eqref{eq:mmse-3} is the maximum entropy
bound on differential entropy expressed in terms of
\begin{equation} 
\var\rve = \lambda_\mrm{LMMSE},
\end{equation}
where $\rve$ is the estimation error
\begin{equation} 
\rve = \left(\rvy_\mrm{r} - \bal_\mrm{LMMSE}\rvby_\mrm{e}\right).
\end{equation}

It remains only to verify that the above inequalities are tight for a
Gaussian distribution.  To see this, note that \eqref{eq:mmse-2} holds
with equality when $\rvbx$ is Gaussian (and thus
$(\rvy_\mrm{r},\rvby_\mrm{e})$ are jointly Gaussian) since in this
case $\rve$ is the (unconstrained) MMSE estimation error and is
therefore independent of the ``data'' $\rvby_\mrm{e}$.  Furthermore,
note that in this case \eqref{eq:mmse-3} holds with equality since the
Gaussian distribution maximizes differential entropy subject to a
variance constraint.

\end{proof}

\section{MISOME Secrecy Capacity Derivation}
\label{sec:MISO}

In this section we derive the MISOME capacity and its high and low SNR
asymptotes. 

\subsection{Proof of Theorem~\protect\ref{thm:MISOMEcap}} 
\label{sec:MISOMEcap-proof}

% \subsection*{Achievability}

Achievability of \eqref{eq:MISOMEcap} follows from evaluating
\eqref{eq:CK} with the particular choices
\begin{equation}
\rvu \sim \CN(0,P),\quad\rvbx = \bpsmax \rvu,
\label{eq:CKParams}
\end{equation} 
where $\bpsmax$ is as defined in Theorem~\ref{thm:MISOMEcap}.  With
this choice of parameters,
\begin{align}
I(\rvu;&\rvy_\mrm{r})-I(\rvu;\rvby_\mrm{e}) \notag\\
&= I(\rvbx;\rvy_\mrm{r})-I(\rvbx;\rvby_\mrm{e}),\label{eq:MarkovCondn}\\
&= \log\left(1 + P|\bhr^\dagger\bpsmax|^2\right) - \log\left(1 + P\|\bHe\bpsmax\|^2\right) \label{eq:CKparamsChoice}\\
&= \log\frac{\bpsmax^\dagger(\bI +
P\bhr\bhr^\dagger)\bpsmax}{\bpsmax^\dagger(\bI + P \bHe^\dagger \bHe
)\bpsmax}\notag\\
&=\log\lamax(\bI + P\bhr\bhr^\dagger, \bI + P \bHe^\dagger
\bHe)\label{eq:variationChar},
\end{align}
where \eqref{eq:MarkovCondn} follows from the fact that $\rvbx$ is a
deterministic function of $\rvu$, \eqref{eq:CKparamsChoice} follows
from the choice of $\rvbx$ and $\rvu$ in \eqref{eq:CKParams}, and
\eqref{eq:variationChar} follows from the variational characterization
of generalized eigenvalues \eqref{eq:GenEigRayleigh}.

%\subsection*{Converse}

We next show a converse---that rates greater than \eqref{eq:MISOMEcap}
are not achievable using our upper bound.  Specifically, we show that
\eqref{eq:MISOMEcap} corresponds to our upper bound expression
\eqref{eq:ub} in Theorem~\ref{thm:ub}.

It suffices to show that a particular choice of $\bph$ that is
admissible (i.e., such that $\bK_\bph\in\cK_\bph$) minimizes
\eqref{eq:ub}.  We can do this by showing that
\begin{equation} 
\max_{\bK_P\in\cK_P}R_+(\bK_P,\bK_\bph)
\label{eq:ub-ph}
\end{equation}
with the chosen $\bph$ corresponds to \eqref{eq:MISOMEcap}.

Since only the first term on the right hand side of
\begin{equation*} 
R_+(\bK_P,\bK_\bph) = I(\rvbx;\rvy_\mrm{r}|\rvby_\mrm{e}) =
h(\rvy_\mrm{r}|\rvby_\mrm{e})-h(\rvz_\mrm{r}|\rvbz_\mrm{e})
\end{equation*}
depends on $\bK_P$, we can restrict our attention to maximizing this
first term with respect to $\bK_P$.

Proceeding, exploiting that all variables are jointly Gaussian, we
express this first term in the form of the optimization
\begin{align}
h(\rvy_\mrm{r} | \rvby_\mrm{e}) &= \min_{\bth \in \compls^{\Ne}}
h( \rvy_\mrm{r} - \bth^\dagger \rvby_\mrm{e}) \label{eq:startNumber}\\
&= \min_{\bth \in \compls^{\Ne}}h( (\bhr - \bHe^\dagger \bth)^\dagger \rvbx + \rvz_\mrm{r} - \bth^\dagger \rvbz_\mrm{e})\notag\\
&= \min_{\bth \in \compls^{\Ne}} 
\log\bigl[ (\bhr - \bHe^\dagger \bth)^\dagger \bK_P
(\bhr - \bHe^\dagger \bth) \notag\\
&\qquad\qquad\qquad\qquad {} + 1 + \|\bth\|^2- 2 \R\{\bth^\dagger
  \bph\} \bigr],  \notag
\end{align}
and bound its maximimum over $\bK_P$ according to
\begin{align}
\max_{\bK_P \in \cK_P}& h(\rvy_\mrm{r} | \rvby_\mrm{e}) \notag\\
&=\max_{\bK_P \in \cK_P} \min_{\bth \in \compls^{\Ne}} 
\log\bigl[ (\bhr - \bHe^\dagger \bth)^\dagger \bK_P
  (\bhr - \bHe^\dagger \bth) \notag\\
&\qquad\qquad\qquad\qquad\qquad{} + 1 + \|\bth\|^2- 2 \R\{\bth^\dagger \bph\} \bigr]\notag\\
&\le  \min_{\bth \in \compls^{\Ne}} \max_{\bK_P \in \cK_P} 
\log\bigl[
(\bhr - \bHe^\dagger \bth)^\dagger \bK_P (\bhr -
  \bHe^\dagger \bth) \notag \\
&\qquad\qquad\qquad\qquad\qquad{} + 1 + \|\bth\|^2- 2
  \R\{\bth^\dagger \bph\} \bigr]\notag\\ 
&= \min_{\bth \in \compls^{\Ne}} 
\log\bigl[ P\|\bhr\!-\!\bHe^\dagger \bth\|^2 
\!+\!1\!+\!\|\bth\|^2\!-\!2\R\{\bth^\dagger \bph\} \bigr],
\label{eq:entropyUB}  
\end{align}
where \eqref{eq:entropyUB} follows by observing that a rank one
$\bK_P$ maximizes the quadratic form $(\bhr -
\bHe^\dagger \bth)^\dagger \bK_P (\bhr -
\bHe^\dagger \bth)$.  

Note that directly verifying that rank one covariance maximizes the term $h(\rvy_\mrm{r} | \rvby_\mrm{e})$ appears difficult.  The above elegant derivation between~\eqref{eq:startNumber} and~\eqref{eq:entropyUB} was suggested to us by Yonina C. Eldar and Ami Wiesel. In the literature, this line of reasoning has been used in deriving an extremal characterization of the Schur complement of a matrix (see e.g., \cite[Chapter 20]{MarshallOlkin},\cite{LiMathias:00}).

We now separately consider the cases $\lamax>1$ and $\lamax\le 1$.

\subsubsection*{Case: $\lamax > 1$}

We show that the choice
\begin{equation} 
\bph = \frac{\bHe\bpsmax}{\bhr^\dagger \bpsmax} 
\label{eq:bph-opt-I}
\end{equation}
in \eqref{eq:ub-ph} yields \eqref{eq:MISOMEcap}, i.e., $\log\lamax$.

We begin by noting that since $\lamax > 1$, the variational
characterization \eqref{eq:GenEigRayleigh} establishes that
$\|\bph\|<1$ and thus $\bK_\bph\in\cK_\bph$ as defined in
\eqref{eq:Kph-def}.

Then, provided that, with $\bph$ as given in \eqref{eq:bph-opt-I}, the right
hand side of \eqref{eq:entropyUB} evaluates to
\begin{multline}
\min_{\bth \in \compls^{\Ne}} 
\log\bigl[ P\|\bhr\!-\!\bHe^\dagger \bth\|^2 
\!+\!1\!+\!\|\bth\|^2\!-\!2\R\{\bth^\dagger \bph\} \bigr] \\
= \log\left(\lamax\cdot(1-\|\bph\|^2)\right),
\label{eq:maxEntClaim}
\end{multline}
we have
\begin{align*}
R_+&\le\max_{\bK_P\in\cK_P}R_+(\bK_P,K_{\bph})\label{eq:specificPhi}\\
&=\max_{\bK_P\in\cK_P}h(\rvy_\mrm{r}|\rvby_\mrm{e}) -
h(\rvz_\mrm{r}|\rvbz_\mrm{e})\\ 
&\le \log(\lamax\cdot(1-\|\bph\|^2)) -\log(1-\|\bph\|^2)\\
&=\log(\lamax),
\end{align*}
i.e., \eqref{eq:MISOMEcap}, as required.  Verifying
\eqref{eq:maxEntClaim} with \eqref{eq:bph-opt-I} is a straightforward
computation, the details of which are provided in
Appendix~\ref{app:maxEntClaim}.

\subsubsection*{Case: $\lamax \le 1$, $\bHe$ full column rank}

We show that the choice
\begin{equation} 
\bph = \bHe (\bHe^\dagger \bHe)^{-1} \bhr
\label{eq:bph-opt-II}
\end{equation}
in \eqref{eq:ub-ph} yields \eqref{eq:MISOMEcap}, i.e., zero.

To verify that $\|\bph\|\le 1$, first note that since $\lamax\le1$,
it follows from \eqref{eq:GenEigRayleigh} that
\begin{equation}
\lamax(\bI + P \bhr\bhr^\dagger, \bI + P \bHe^\dagger
\bHe)\le 1 \Leftrightarrow \lamax(\bhr\bhr^\dagger,
\bHe^\dagger \bHe)\le 1,
\label{eq:lamax-special-a}
\end{equation}
so that for any choice of $\bps$, 
\begin{equation} 
\bps^\dagger\bhr\bhr^\dagger\bps \le
\bps^\dagger\bHe^\dagger\bHe\bps.
\label{eq:lamax-special-b}
\end{equation}
Choosing $\bps=(\bHe^\dagger\bHe)^{-1}\bhr$ in
\eqref{eq:lamax-special-b} yields $\|\bph\|^2\le\|\bph\|$, i.e.,
$\|\bph\|\le1$, as required.

Next, note that \eqref{eq:entropyUB} is further upper bounded by
choosing any particular choice of $\bth$.  Choosing $\bth=\bph$ yields
\begin{equation}
R_+ \le \log\left(
\frac{P\|\bhr-\bHe^\dagger\bph\|^2}{1-\|\bph\|^2}+1
\right)
\end{equation}
which with the choice \eqref{eq:bph-opt-II} for $\bph$ is zero.

\subsubsection*{Case: $\lamax \le 1$, $\bHe$ not full column rank}

Consider a new MISOME channel with $\Nt'<\Nt$ transmit antennas, where
$\Nt'$ is the column rank of $\bHe$, where the intended
receiver and eavesdropper channel gains are given by
\begin{equation} 
\bgr = \bQ^\dagger \bhr, \qquad
\bGe = \bHe \bQ,
\label{eq:gG-def}
\end{equation}
and where $\bQ$ is a matrix whose columns constitute an orthogonal
basis for the column space of $\bHe^\dagger$, so that in this new
channel $\bGe$ has full rank.

Then provided the new channel \eqref{eq:gG-def} has the same capacity
as the original channel, it follows by the analysis of the previous
case that the capacity of both channels is zero.  Thus it remains only
to show the following.
\begin{claim}
The MISOME channel $(\bgr,\bGe)$ corresponding to \eqref{eq:gG-def}
has the same secrecy capacity as that corresponding to $(\bhr,\bHe)$.
\end{claim}

\begin{proof}
First we show that the new channel capacity is no larger than the
original one.  In particular, we have
\begin{align} 
\lamax(\bI + P\bgr\bgr^\dagger,
&\bI + P\bGe^\dagger \bGe) \notag\\
&= \max_{\{\bps':\|\bps'\|=1\}}\left\{\frac{1 + P|\bgr^\dagger\bps'|^2}{1 +
P\|\bGe\bps'\|^2}\right\} \label{eq:glamax}\\ 
&=\max_{\{\bps':\|\bps'\|=1\}} \frac{1 + P|\bhr^\dagger \bQ\bps'|^2}{1 +
P\|\bHe\bQ\bps'\|^2} \label{eq:glamax-exp}\\
&=\max_{\{\bps:\bps=\bQ\bps', \|\bps\|=1\}} \frac{1 +
  P|\bhr^\dagger \bps|^2}{1 + P\|\bHe\bps\|^2} \label{eq:hlamax-exp-a}\\
&\le \max_{\{\bps:\|\bps\|=1\}}\left\{\frac{1 + P|\bhr^\dagger\bps|^2}{1 +
P\|\bHe\bps\|^2}\right\} \label{eq:hlamax-exp}\\
&= \lamax(\bI + P\bhr\bhr^\dagger,\bI + P\bHe^\dagger
\bHe), \label{eq:hlamax}
\end{align}
where to obtain \eqref{eq:glamax} we have used
\eqref{eq:GenEigRayleigh} for the new channel, to obtain
\eqref{eq:glamax-exp} we have used \eqref{eq:gG-def}, to obtain
\eqref{eq:hlamax-exp-a} we have used that $\bQ^\dagger\bQ=\bI$, to obtain
\eqref{eq:hlamax-exp} we have used that we are maximizing over a
larger set, and to obtain \eqref{eq:hlamax} we have used
\eqref{eq:GenEigRayleigh} for the original channel.   
Thus,
\begin{multline}
\left\{\lamax(\bI + P\bgr\bgr^\dagger,
\bI + P\bGe^\dagger \bGe)\right\}^+ \\
\le \left\{\lamax(\bI + P\bhr\bhr^\dagger,
\bI + P\bHe^\dagger \bHe)\right\}^+,
\label{eq:gG-le}
\end{multline}

Next, we show the new channel capacity is no smaller than the original
one.   To begin, note that 
\begin{equation} 
\Null(\bHe) \subseteq \Null(\bhr^\dagger),
\label{eq:null-nesting}
\end{equation}
since if $\Null(\bHe) \nsubseteq \Null(\bhr^\dagger)$,
then $\lamax(\bhr\bhr^\dagger, \bHe^\dagger
\bHe)=\infty$, which would violate \eqref{eq:lamax-special-a}.

Proceeding, every $\bx \in \compls^{\Nt}$ can we written as
\begin{equation} 
\bx = \bQ\bx' +  \tilde{\bx},
\label{eq:x-decomp}
\end{equation}
where $\bHe\tilde{\bx} = \mathbf{0}$ and thus, via
\eqref{eq:null-nesting}, $\bhr^\dagger\tilde{\bx} = 0$ as well.
Hence, we have that $\bhr^\dagger \bx = \bgr^\dagger
\bx'$, $\bHe\bx = \bGe\bx'$, and $\|\bx'\|^2 \le
\|\bx\|^2$, so any rate achieved by $p_{\rvbx}$ on the channel
$(\bhr,\bHe)$ is also achieved by $p_{\rvbx'}$ on the
channel $(\bgr,\bGe)$, with $p_{\rvbx'}$ derived from 
$p_\rvbx$ via \eqref{eq:x-decomp}, whence
\begin{multline}
\left\{\lamax(\bI + P\bgr\bgr^\dagger,
\bI + P\bGe^\dagger \bGe)\right\}^+ \\
\ge \left\{\lamax(\bI + P\bhr\bhr^\dagger,
\bI + P\bHe^\dagger \bHe)\right\}^+.
\label{eq:gG-ge}
\end{multline}
Combining \eqref{eq:gG-ge} and \eqref{eq:gG-le} establishes our
claim.
\end{proof}

\subsection{Proof of~Corollary~\protect\ref{corol:high-SNR}} 
\label{sec:high-SNR-proof}

We restrict our attention to the case $\lamax>1$ where
the capacity is nonzero.  In this case, since, via \eqref{eq:GenEigRayleigh}, 
\begin{equation} 
\lamax(\bI+P\bhr\bhr^\dagger,\bI+P\bHe^\dagger \bHe)
= \frac{1 + P|\bhr^\dagger\bpsmax(P)|^2}{1+ P\|\bHe\bpsmax(P)\|^2} >1,
\label{eq:varcharcap} 
\end{equation}
where
\begin{equation} 
\bpsmax(P)\defeq \argmax_{\{\bps: \|\bps\|=1\}} \frac{1 +
P|\bhr^\dagger\bps|^2}{1+ P\|\bHe\bps\|^2},
\label{eq:bpsmax-def}
\end{equation}
we have
\begin{equation} 
|\bhr^\dagger \bpsmax(P)| > \|\bHe\bpsmax(P)\|
\label{eq:poscapcond}
\end{equation}
for all $P>0$.

To obtain an upper bound note that, for all $P>0$,
\begin{align}
\lamax(\bI+P\bhr\bhr^\dagger,
&\bI+P\bHe^\dagger \bHe) \notag\\
&\le\frac{|\bhr^\dagger\bpsmax(P)|^2}{\|\bHe\bpsmax(P)\|^2}
  \label{eq:varcharbnd-a}\\ 
&\le
\lamax(\bhr\bhr^\dagger, \bHe^\dagger
\bHe), \label{eq:UpperLim1}
\end{align}
where \eqref{eq:varcharbnd-a} follows from the Rayleigh quotient
expansion \eqref{eq:varcharcap} and the fact that, due to
\eqref{eq:poscapcond}, the right hand side of \eqref{eq:varcharcap} is
increasing in $P$, and where \eqref{eq:UpperLim1} follows from
\eqref{eq:GenEigRayleigh}.  Thus, since the right hand side of
\eqref{eq:UpperLim1} is independent of $P$ we have
\begin{equation}
\lim_{P\rightarrow\infty}\lamax(\bI+P\bhr\bhr^\dagger,
\bI+P\bHe^\dagger \bHe) \le
\lamax(\bhr\bhr^\dagger, \bHe^\dagger \bHe).
\label{eq:UpperLim}
\end{equation}

Next, defining 
\begin{equation} 
\bpsmax(\infty) \defeq \argmax_{\bps}
\frac{|\bhr^\dagger\bps|^2}{\|\bHe\bps\|^2}, 
\label{eq:bpsinfty-def}
\end{equation}
we have the lower bound
\begin{align}
\lim_{P\rightarrow\infty} \lamax(\bI+P\bhr\bhr^\dagger,
&\bI+P\bHe^\dagger \bHe) \notag\\
&\ge \lim_{P\rightarrow\infty}\frac{1/P +
|\bhr^\dagger\bpsmax(\infty)|^2}{1/P+
  \|\bHe\bpsmax(\infty)\|^2} \label{eq:varcharbnd-b}\\
&=\lamax(\bhr\bhr^\dagger, \bHe^\dagger \bHe)
\label{eq:LowerLim}
\end{align}
where \eqref{eq:varcharbnd-b} follows from \eqref{eq:GenEigRayleigh}
and \eqref{eq:LowerLim} follows from \eqref{eq:bpsinfty-def}.

Since \eqref{eq:UpperLim} and \eqref{eq:LowerLim} coincide we obtain
\eqref{eq:cap-high-SNR}.  Thus, to obtain the remainder of
\eqref{eq:high-SNR-a} we need only verify the following.
\begin{claim}
The high SNR capacity is finite, i.e.,
$\lamax(\bhr\bhr^\dagger,\bHe^\dagger\bHe)<\infty$, when $\bHe^\perp
\bhr=\mathbf{0}$.
\end{claim}

\begin{proof}
We argue by contradiction.  Suppose
$\lamax(\bhr\bhr^\dagger,\bHe^\dagger\bHe)=\infty$.  Then there must
exist a sequence $\bps_k$ such that $\|\bHe\bps_k\|>0$ for each
$k=1,2,\dots$, but $\|\bHe\bps_k\|\rightarrow0$ as
$k\rightarrow\infty$.  But then the hypothesis cannot be true,
because, as we now show, $|\bhr^\dagger\bps|^2/\|\bHe\bps\|^2$, when
viewed as a function of $\bps$, is bounded whenever the denominator is
nonzero.

Let $\bps$ be any vector such that $\|\bHe\bps\|\defeq\delta>0$.  It
suffices to show that
\begin{equation} 
\frac{|\bhr^\dagger\bps|^2}{\|\bHe\bps\|^2} \le
\frac{\|\bhr\|^2}{\sigma^2},
\label{eq:limcapbnd}
\end{equation}
where $\sigma^2$ is the smallest \emph{nonzero} singular value of $\bHe$.

To verify \eqref{eq:limcapbnd}, we first express $\bps$ in the form
\begin{equation} 
\bps = c\bps' + d\tilde{\bps},
\label{eq:bps-decomp}
\end{equation}
where $\bps'$ and $\tilde{\bps}$ are unit vectors, $c$ and
$d$ are real and nonnegative, $d\tilde{\bps}$ is the
projection of $\bps$ onto the null space of $\bHe$, and $c\bps'$
is the projection of $\bps$ onto the orthogonal complement of this null
space.

Next, we note that 
$\delta=\|\bHe\bps\| = c\|\bHe\bps'\| \ge c\sigma$, whence
\begin{equation} 
c \le \frac{\delta}{\sigma}.
\label{eq:delta-rel}
\end{equation}

But since $\bHe^\perp\bhr=0$ it follows that
$\bhr^\dagger\tilde{\bps}=0$, so
\begin{equation} 
|\bhr^\dagger\bps|^2 = c^2|\bhr^\dagger\bps'|^2 \le c^2\|\bhr\|^2 \le
\frac{\delta^2}{\sigma^2}\|\bhr\|^2,
\label{eq:hr-ip-bps}
\end{equation}
where the first inequality follows from the Cauchy-Schwarz inequality,
and the second inequality is a simple substitution from
\eqref{eq:delta-rel}.   Dividing through by $\|\bHe\bps\|^2=\delta^2$ in
\eqref{eq:hr-ip-bps} yields \eqref{eq:limcapbnd}.

\end{proof}

We now develop \eqref{eq:high-SNR-b} for the case where $\bHe^\perp
\bhr \neq \mathbf{0}$.  

First, defining
\begin{equation}
\cS_\infty = \{\bps : \|\bps\|=1, \|\bHe\bps\|=0\}
\end{equation}
we obtain the lower bound
\begin{align}
&\frac{1}{P}\lamax(\bI + P\bhr\bhr^\dagger, \bI + P\bHe^\dagger \bHe) \notag\\
&\ge\max_{\bps\in \cS_\infty} \frac{1/P +
|\bhr^\dagger\bps|^2}{1+P\|\bHe\bps\|^2} \notag\\
&=\max_{\bps\in \cS_\infty} {\frac{1}{P} + |\bhr^\dagger\bps|^2}\\
&= {\frac{1}{P} + \|\bHe^\perp\bhr\|^2}, \label{eq:HrBound1}
\end{align}
where to obtain~\eqref{eq:HrBound1} we have used,
\begin{equation} 
\max_{\{\bps: \|\bps\|=1, \bHe\bps=0\}} 
\left|\bhr^\dagger\bps\right|^2 = \left\|\bHe^\perp \bhr\right\|^2.
\label{eq:bHe-perp-char}
\end{equation}

Next we develop an upper bound.  We first establish the following.
\begin{claim}
If $\bHe^\perp \bhr \neq \mathbf{0}$ then there is a function $\eps(P)$
such that $\eps(P) \rightarrow 0$ as $P\rightarrow \infty$, and
\begin{equation*} 
\|\bHe \bpsmax(P)\|\le \eps(P).
\end{equation*}
\label{claim:domain}
\end{claim}

\begin{proof}
We have
\begin{align}
\frac{1 + P \|\bhr\|^2}{1+P\|\bHe\bpsmax(P)\|^2}  
&\ge
\frac{1 + P |\bhr^\dagger\bpsmax(P)|^2}{1+P\|\bHe\bpsmax(P)\|^2}
\label{eq:lam-lt1} \\ 
&\ge \max_{\{\bps: \bHe\bps=\mathbf{0}, \|\bps\|=1\}} 
\frac{1 + P |\bhr^\dagger\bps|^2}{1+P\|\bHe\bps\|^2} \label{eq:reduce-set} \\ 
&= \max_{\{\bps: \bHe\bps=\mathbf{0}, \|\bps\|=1\}} 
\left(1 + P |\bhr^\dagger\bps|^2 \right) \notag \\
&= 1 + P\|\bHe^\perp \bhr\|^2 \label{eq:perp-version}
%&\ge P\|\bHe^\perp \bhr\|^2.  \label{eq:foobar}
\end{align}
where to obtain \eqref{eq:lam-lt1} we have used the Cauchy-Schwarz
inequality $|\bhr^\dagger\bpsmax(P)|^2 \le \|\bhr\|^2$, to obtain
\eqref{eq:reduce-set} we have used \eqref{eq:bpsmax-def}, and to
obtain \eqref{eq:perp-version} we have used~\eqref{eq:bHe-perp-char}.

Rearranging \eqref{eq:perp-version} then gives
\begin{equation*} 
\|\bHe\bpsmax(P)\|^2\le
\frac{1}{P}\left(\frac{1+P\|\bhr\|^2}{1+P\|\bHe^\perp
  \bhr\|^2}-1\right)\defeq \eps^2(P).
\end{equation*}
as desired.
\end{proof}

Thus with $\cS_P = \{\bps : \|\bps\|=1, \|\bHe\bps\|\le \eps(P)\}$ we have
\begin{align}
&\frac{1}{P}\lamax(\bI + P\bhr\bhr^\dagger, \bI + P\bHe^\dagger \bHe) \notag\\
&=\max_{\bps\in \cS_P} \frac{1/P +
|\bhr^\dagger\bps|^2}{1+P\|\bHe\bps\|^2} \label{eq:optDomain}\\
&\le \max_{\bps\in \cS_P} {\frac{1}{P} +
|\bhr^\dagger\bps|^2} \label{eq:HrBound2},
\end{align}
where \eqref{eq:optDomain} follows from \eqref{eq:GenEigRayleigh} and
Claim~\ref{claim:domain} that the maximizing $\bpsmax$ lies in
$\cS_P$.

Now, as we will show, 
\begin{equation} 
\max_{\bps\in \cS_P} |\bhr^\dagger\bps|^2 \le \|\bHe^\perp
\bhr\|^2 + \frac{\eps^2(P)}{\sigma^2}\|\bhr\|^2.
\label{eq:continuity1}
\end{equation}
so using \eqref{eq:continuity1} in \eqref{eq:HrBound2} we obtain
\begin{equation}
\begin{split}
\frac{1}{P}\lamax(\bI + &P\bhr\bhr^\dagger, \bI + P\bHe^\dagger \bHe) \\
&\le \|\bHe^\perp\bhr\|^2 + \frac{\eps^2(P)}{\sigma^2}\|\bhr\|^2 + \frac{1}{P}
\end{split}
\label{eq:geigUB}
\end{equation}
Finally, combining \eqref{eq:geigUB} and \eqref{eq:HrBound1} we obtain
\begin{equation*}
\lim_{P\rightarrow\infty}
\frac{1}{P}\lamax(\bI + P\bhr\bhr^\dagger, \bI + P\bHe^\dagger \bHe) 
= \|\bHe^\perp\bhr\|^2,
\end{equation*}
whence \eqref{eq:high-SNR-b}.

Thus, it remains only to verify \eqref{eq:continuity1}, which we do now.

%Let $\bpsmax'(P)$ be the maximizing vector.  Note that since
%$\bpsmax'(P)\in \cS_P$, we have that $\|\bpsmax'(P)\|=1$ and
%$\|\bHe\bpsmax'(P)\|\le \eps(P)$.  

We start by expressing $\bps\in\cS_P$ in the form [\cf\ \eqref{eq:bps-decomp}]
\begin{equation} 
\bps = c\bps' + d\tilde{\bps},
\label{eq:bps-decomp-again}
\end{equation}
where $\bps'$ and $\tilde{\bps}$ are unit vectors, $c,d$ are
real valued scalars in $[0,1]$, $d\tilde{\bps}$ is the projection of $\bps$ onto
the null space of $\bHe$, and $c\bps'$ is the projection of $\bps$
onto the orthogonal complement of this null space.  

With these definitions we have,
\begin{equation} 
\eps(P) \ge \|\bHe \bps\| = c \|\bHe\bps'\| \ge c\sigma
\label{eq:c2-bnd}
\end{equation}
since $\bHe\tilde{\bps}=\mathbf{0}$ and $\|\bHe\bps'\|\ge \sigma$.

Finally, 
\begin{align}
|\bhr^\dagger\bps|^2 
&= |d\bhr^\dagger\tilde{\bps} + c \bhr^\dagger \bps'|^2 \label{eq:bps-sub}\\
&= d^2 |\bhr^\dagger \tilde{\bps}|^2 + c^2 |\bhr^\dagger \bps'|^2\label{eq:exploit-orthog}\\
&\le |\bhr^\dagger \tilde{\bps}|^2 +
\frac{\eps(P)^2}{\sigma^2}|\bhr^\dagger\bps'|^2\label{eq:c2-sub}\\ 
&\le |\bhr^\dagger \tilde{\bps}|^2 + \frac{\eps(P)^2}{\sigma^2}\|\bhr\|^2\\
&\le \|\bHe^\perp \bhr\|^2 + \frac{\eps(P)^2}{\sigma^2}\|\bhr\|^2,\label{eq:perp-form}
\end{align}
where \eqref{eq:bps-sub} follows from substituting
\eqref{eq:bps-decomp-again}, \eqref{eq:exploit-orthog} follows from
the fact that $\bps'$ and $\tilde{\bps}$ are orthogonal,
\eqref{eq:c2-sub} follows from using \eqref{eq:c2-bnd} to bound $c^2$,
and \eqref{eq:perp-form} follows from the fact that
$\bHe\tilde{\bps}=\mathbf{0}$ and \eqref{eq:bHe-perp-char}.

\subsection{Proof of~Corollary~\protect\ref{corol:low-SNR} } 
\label{sec:low-SNR-proof}

We consider the limit $P\rightarrow 0$.   In the following steps, the
order notation $\cO(P)$ means that $\cO(P)/P \rightarrow 0$ as $P
\rightarrow 0$.
\begin{align}
\lamax(&\bI + P\bhr \bhr^\dagger, \bI + P\bHe^\dagger
\bHe)\\
&= \lamax\left((\bI + P\bHe^\dagger \bHe)^{-1}(\bI +
P\bhr\bhr^\dagger)\right)\label{eq:geigDefn}\\
&= \lamax\left((\bI - P\bHe^\dagger \bHe + \cO(P))(\bI +
P\bhr\bhr^\dagger)\right)\label{eq:TaylorExp}\\
&= \lamax\left((\bI - P\bHe^\dagger \bHe )(\bI +
P\bhr\bhr^\dagger)\right) + \cO(P) \label{eq:EigCont}\\
&= \lamax\left(\bI + P(\bhr\bhr^\dagger - \bHe^\dagger
\bHe)\right)+\cO(P)\label{eq:EigCont2}\\
&= 1 + P\lamax(\bhr\bhr^\dagger - \bHe^\dagger \bHe) +
\cO(P)\label{eq:EigProp},
\end{align}
where \eqref{eq:geigDefn} follows from the definition of generalized
eigenvalue, \eqref{eq:TaylorExp} follows from the Taylor series
expansion of $(\bI+P\bHe^\dagger \bHe)^{-1}$, where we have assumed that
$P$ is sufficiently small so that all eigenvalues of $P\bHe^\dagger
\bHe$ are less than unity, \eqref{eq:EigCont} and \eqref{eq:EigCont2}
follow from the continuity of the eigenvalue function in its
arguments and \eqref{eq:EigProp} follows from the property of
eigenvalue function that $\lambda(\bI + \bA) = 1 +\lambda(\bA)$.

In turn, we have,
\begin{align}
\frac{C(P)}{P}
&=\frac{\log(1 + P\lamax(\bhr\bhr^\dagger - \bHe^\dagger
\bHe) + \cO(P))}{P} \label{eq:ts-a}\\
&= \frac{\lamax(\bhr\bhr^\dagger - \bHe^\dagger \bHe)}{\ln 2} +
\frac{\cO(P)}{P}, \label{eq:ts-b}
\end{align}
where to obtain \eqref{eq:ts-a} we have used \eqref{eq:EigProp} in
\eqref{eq:MISOMEcap}, and to obtain \eqref{eq:ts-b} we have used
Taylor Series expansion of the $\ln(\cdot)$ function.  

Finally, taking the limit $P\rightarrow 0$ in \eqref{eq:ts-b} yields
\eqref{eq:low-SNR} as desired. 

\section{Masked Beamforming Scheme Analysis}
\label{sec:AN} 

\iffalse
In the capacity-achieving scheme in
Theorem~\ref{thm:MISOMEcap}, the optimal transmit direction is
the generalized eigenvector corresponding to $\lamax(\bI
+ P \bhr\bhr^\dagger, \bI + P \bHe^\dagger \bHe)$.  The knowledge of
both $\bhr$ and $\bHe$ have been used in choosing the transmit
directions.  In this section we consider a suboptimal scheme that
only uses the knowledge of $\bhr$ in choosing the transmit
directions (the knowledge of $\bHe$ is still necessary in choosing
the rate).   We note that this scheme was proposed
in~\cite{NegiGoel05,GoelNegi05}, but the optimality claims
presented here appear to be new.

\subsection{Description of the Scheme}

\fi

From Csisz{\'a}r-K{\"o}rner \cite{csiszarKorner:78}, secrecy rate
$R=I(\rvu;\rvy_\mrm{r})-I(\rvu;\rvby_\mrm{e})$ is achievable for any
choice of $p_\rvu$ and $p_{\rvbx|\rvu}$ that satisfy the power
constraint $E[|\rvx|^2]\le P$.  While a capacity-achieving scheme
corresponds to maximizing this rate over the choice of $p_\rvu$ and
$p_{\rvbx|\rvu}$ (\cf\ \eqref{eq:CK}), the masked beamforming scheme
corresponds to different (suboptimal) choice of these distributions.
In particular, we choose
\begin{equation} 
p_\rvu = \CN(0,\tilde{P}) \quad\text{and}\quad p_{\rvbx|\rvu} =
\CN(\rvu\tbhr,\tilde{P}(\bI-\tbhr\tbhr^\dagger), 
\label{eq:MB-dists}
\end{equation}
where we have chosen the convenient normalizations
\begin{equation} 
\tilde{P}= \frac{P}{\Nt}
\label{eq:tP-def}
\end{equation}
and 
\begin{equation} 
\tbhr = \frac{\bhr}{\|\bhr\|}.
\label{eq:tbhr-def}
\end{equation}

In this form, the secrecy rate of masked beamforming is readily
obtained, as we now show

\subsection{Proof of Proposition~\protect\ref{prop:RMB}}
\label{sec:RMB-proof}

With $p_\rvu$ and $p_{\rvbx|\rvu}$ as in \eqref{eq:MB-dists}, we
evaluate \eqref{eq:CK}.  To this end, first we have
\begin{equation}
I(\rvu;\rvy_\mrm{r}) = \log(1 + \tilde{P}\|\bhr\|^2)
\label{eq:Iuyr}
\end{equation}
Then, to evaluate $I(\rvu;\rvby_\mrm{e})$, note that
\begin{align*}
h(\rvby_\mrm{e}) &=\log\det(\bI + \tilde{P} \bHe \bHe^\dagger)\\
h(\rvby_\mrm{e}|\rvu) 
&= \log\det(\bI + \tilde{P}\bHe(\bI - \tbhr\tbhr^\dagger)\bHe^\dagger)
\end{align*}
so
\begin{align}
I(\rvu;&\rvby_\mrm{e})\notag\\
&=h(\rvby_\mrm{e})-h(\rvby_\mrm{e}|\rvu)\notag\\
&=\log\det(\bI\!+\!\tilde{P} \bHe \bHe^\dagger)\!-\!\log\det(\bI + \tilde{P}\bHe(\bI\!-\!
\tbhr\tbhr^\dagger)\bHe^\dagger)\notag\\
&=\log\det(\bI\!+\!\tilde{P} \bHe^\dagger \bHe)\!-\!\log\det(\bI + \tilde{P}(\bI\!-\!
\tbhr\tbhr^\dagger)\bHe^\dagger \bHe)\notag\\
&=\log\det(\bI + \tilde{P} \bHe^\dagger \bHe)\notag\\
&\quad\qquad{}- \log\det(\bI + \tilde{P} \bHe^\dagger \bHe
\!-\!\tilde{P}\tbhr\tbhr^\dagger \bHe^\dagger \bHe)\notag\\
&=-\log\det\left(\bI
-\tilde{P}\tbhr\tbhr^\dagger \bHe^\dagger \bHe(\bI + \tilde{P}
\bHe^\dagger \bHe)^{-1}\right) \notag\\
&=-\log\left(1
-\tilde{P}\tbhr^\dagger \bHe^\dagger \bHe(\bI + \tilde{P} \bHe^\dagger
\bHe)^{-1}\tbhr\right) \notag \\ 
&=-\log\left(\tbhr^\dagger(\bI + \tilde{P} \bHe^\dagger \bHe)^{-1}\tbhr\right),
\label{eq:Iuye}
\end{align}
where we have repeatedly used the matrix identity
$\det(\bI+\bA\bB)=\det(\bI+\bB\bA)$ valid for any $\bA$ and $\bB$ with
compatible dimensions.

Thus, combining \eqref{eq:Iuyr} and \eqref{eq:Iuye} we obtain
\eqref{eq:RMB} as desired:
\begin{align*}
R_\mrm{MB}&(P) \\
&= I(\rvu;\rvy_\mrm{r})-I(\rvu;\rvby_\mrm{e})\\&= \log(1 +
\tilde{P}\|\bhr\|^2)+\log(\tbhr^\dagger(\bI + \tilde{P}
\bHe^\dagger \bHe)^{-1}\tbhr)\\
&= \log\left(1\!+\!
\frac{1}{\tilde{P}\|\bhr\|^2}\right)+\log(\tilde{P}\bhr^\dagger(\bI
+ \tilde{P} \bHe^\dagger \bHe)^{-1}\bhr)\\
&=\log\left(1\! +\!
\frac{1}{\tilde{P}\|\bhr\|^2}\right)+\log(\lamax(\tilde{P}\bhr\bhr^\dagger,
\bI\!+\!\tilde{P} \bHe^\dagger \bHe)),
\end{align*}
where to obtain the last equality we have used the special
form \eqref{eq:lamax-rankone} for the largest generalized eigenvalue.

\subsection{Proof of Theorem~\ref{thm:RMBloss}}
\label{sec:RMBloss-proof}

First, from Theorem~\ref{thm:MISOMEcap} and
Proposition~\ref{prop:RMB} we have, with again
$\tilde{P}$ as in \eqref{eq:tP-def} for convenience,
\begin{equation}
C\left(\frac{P}{\Nt}\right)-R_\mrm{MB}(P) \le
\log\frac{\lamax(\bI + \tilde{P}\bhr\bhr^\dagger, \bI +
\tilde{P}\bHe^\dagger \bHe)}{\lamax(\tilde{P}\bhr\bhr^\dagger,\bI +
\tilde{P}\bHe^\dagger \bHe)}.
\label{eq:C-R-exp}
\end{equation}

Next, with $\bpsmax$ denoting the generalized eigenvector corresponding
to $\lamax(\bI + \tilde{P}\bhr\bhr^\dagger, \bI +
\tilde{P}\bHe^\dagger \bHe)$, we have
\begin{align}
\lamax(\bI + \tilde{P}\bhr\bhr^\dagger, \bI + \tilde{P}\bHe^\dagger
\bHe) &=
\frac{1 + \tilde{P} |\bhr^\dagger\bpsmax|^2}{1 + \tilde{P}
  \|\bHe\bpsmax\|^2} \label{eq:lamax-C} \\
\lamax(\tilde{P}\bhr\bhr^\dagger, \bI + \tilde{P}\bHe^\dagger \bHe)
&\ge
\frac{\tilde{P} |\bhr^\dagger\bpsmax|^2}{1 + \tilde{P}
  \|\bHe\bpsmax\|^2} \label{eq:lamax-R} \\
\end{align}

Finally, substituting \eqref{eq:lamax-C} and \eqref{eq:lamax-R} into
\eqref{eq:C-R-exp}, we obtain 
\begin{equation}
0\le C\left(\frac{P}{\Nt}\right)-R_\mrm{MB}(P) \le \log\left(1 +
\frac{\Nt}{P|\bhr^\dagger\bpsmax|^2}\right),
\end{equation}
the right hand side of which approaches zero as $P\rightarrow\infty$,
whence \eqref{eq:RMBloss} as desired.

\section{Scaling Laws Development}
\label{sec:scaling}

We begin by summarizing a few well-known results from random matrix
theory that will be useful in our scaling laws; for further details,
see, e.g., \cite{verduTulino:05}.

\subsection{Some Random Matrix Properties}

Three basic facts will suffice for our purposes.

\begin{fact}
\label{fact:QuadraticForm}
Suppose that $\rvbv$
is a random length-$n$ complex vector with independent,
zero-mean, variance-$1/n$ elements, and that $\rvbB$ is a
random $n\times n$ complex positive semidefinite matrix distributed
independently of $\rvbv$.  Then if the spectrum of $\rvbB$ converges
we have
\begin{equation}
\lim_{n\rightarrow\infty}
\rvbv^\dagger(\bI + \g \rvbB)^{-1}\rvbv \stackrel{\text{a.s.}}{=}
\eta_\rvbB(\g), 
\label{eq:QuadraticForm} 
\end{equation}
where $\eta_\rvbB(\g)$ is the $\eta$-transform \cite{verduTulino:05}
of the matrix $\rvbB$.
\end{fact}

Of particular interest to us is the $\eta$-transform of a special
class of matrices below.
\begin{fact}
Suppose that $\rvbH\in\compls^{K\times N}$ is random matrix whose
entries are i.i.d.\ with variance $1/N$.  As $K,N\rightarrow\infty$
with the ratio $K/N\defeq\beta$ fixed, the $\eta$-transform of
$\rvbB=\rvbH^\dagger \rvbH$ is given by
\begin{equation}
\eta_{\rvbH^\dagger \rvbH}(\g) = 
\frac{\mpf(\g,\beta)}{\g},
\label{eq:etaTrans}
\end{equation}
where  $\mpf(\cdot,\cdot)$ is as defined in \eqref{eq:mpf}.
\label{fact:etaTrans}
\end{fact}

The distribution of generalized eigenvalues of the pair
$(\rvbhr\rvbhr^\dagger, \rvbHe^\dagger
\rvbHe)$ is also known~\cite{KangAlouini03,Muirhead}.  For our
purposes, the following is sufficient.
\begin{fact}
\label{fact:geig-dist}
Suppose that $\rvbhr$ and $\rvbHe$ have i.i.d.\ $\CN(0,1)$ entries,
and $\Ne > \Nt$.  Then
\begin{equation}
\lamax(\rvbhr\rvbhr^\dagger, \rvbHe^\dagger \rvbHe) 
\sim
\frac{2\Nt}{2\Ne-2\Nt +1}F_{2\Nt,2\Ne-2\Nt+1},
\label{eq:FStat} 
\end{equation}
where $F_{2\Nt,2\Ne-2\Nt+1}$ is the F-distribution with $2\Nt$ and
$2\Ne-2\Nt +1$ degrees of freedom, i.e.,
\begin{equation}
F_{2\Nt,2\Ne-2\Nt+1} \stackrel{\mrm{d}}{=}
\frac{\rvv_1/(2\Nt)}{\rvv_2/(2\Ne-2\Nt+1)},
\label{eq:chiSquareRelation}
\end{equation}
where $\stackrel{\mrm{d}}{=}$ denote equality in distribution, and 
where $\rvv_1$ and $\rvv_2$ are independent chi-squared random
variables with $2\Nt$ and $2\Ne-2\Nt+1$ degrees of freedom, respectively.
\end{fact}

Using Fact~\ref{fact:geig-dist} it follows that with $\beta=\Ne/\Nt$ fixed,
\begin{equation}
\lim_{\Nt\rightarrow\infty}
\lamax(\rvbhr\rvbhr^\dagger,
\rvbHe^\dagger \rvbHe) \stackrel{\text{a.s.}}{=}
\frac{1}{\beta-1},\quad\text{when $\beta > 1$}. 
\label{eq:asympF}
\end{equation}
Indeed, from the strong law of large numbers we have that the random variables
$\rvv_1$ and $\rvv_2$ in \eqref{eq:chiSquareRelation} satisfy, for $\beta>1$,
\begin{equation}
\lim_{\Nt\rightarrow\infty} {\frac{\rvv_1}{2\Nt}}
\stackrel{\text{a.s.}}{=} 1,\quad\text{and}\quad
\lim_{\Nt\rightarrow\infty} \frac{\rvv_2}{2\Nt(\beta-1)+1}
\stackrel{\text{a.s.}}{=}1
\label{eq:chi-sq-lim}
\end{equation}
Combining \eqref{eq:chi-sq-lim} with \eqref{eq:chiSquareRelation}
yields \eqref{eq:asympF}.

\subsection{Proof of Theorem~\ref{thm:scaling}}
\label{sec:lb-scaling}

First, from Theorem~\ref{thm:MISOMEcap} we have that
\begin{align}
C(P,\Nt,\Ne) &= \left\{\log\lamax(\bI+P\rvbhr\rvbhr^\dagger, \bI +
P\bHe^\dagger \rvbHe) \right\}^+\notag\\
&\ge \left\{ \log\lamax(P\rvbhr\rvbhr^\dagger, \bI + P\bHe^\dagger
\rvbHe) \right\}^+ \notag\\ 
&= \left\{ \log\left( P\rvbhr^\dagger(\bI + P\bHe^\dagger
\rvbHe)^{-1}\rvbhr \right) \right\}^+, 
\label{eq:MisoSC1}
\end{align}
where \eqref{eq:MisoSC1} follows from the quadratic form
representation \eqref{eq:lamax-rankone} of the generalized eigenvalue.

Rewriting \eqref{eq:MisoSC1} using the notation
\begin{equation} 
\trvbhr = \frac{1}{\sqrt{\Nt}}\rvbhr, \quad\text{and}\quad
\trvbHe = \frac{1}{\sqrt{\Nt}}\rvbHe,
\end{equation}
we then obtain \eqref{eq:MarcenkoPastur} as desired:
\begin{align} 
\tilde{C}(\g,\beta) &= C(\g/\Nt,\Nt,\beta\Nt) \notag\\
&\ge \left\{ \log\left(\g\trvbhr(\bI + \g \trvbHe^\dagger
\trvbHe)^{-1}\trvbhr \right) \right\}^+\notag\\
&\stackrel{\text{a.s.}}{\longrightarrow}
\left\{ \log\xi(\g,\beta) \right\}^+\quad\text{as $\Nt\rightarrow\infty$,}
\label{eq:MISOSC3}
\end{align}
where to obtain \eqref{eq:MISOSC3} we have applied
\eqref{eq:QuadraticForm} and \eqref{eq:etaTrans}.

The derivation of the scaling law \eqref{eq:MarcenkoPastur-R} for the
masked beamforming scheme is analogous.  Indeed, from
Proposition~\ref{prop:RMB} we have
\begin{align*}
R_\mrm{MB}(\g,\Nt,\beta\Nt) &\ge\left\{\log\lamax(\g\trvbhr\trvbhr^\dagger, \bI + \g\trvbHe^\dagger \trvbHe)\right\}^+ \\
&= \left\{ \log \left( \g\trvbhr^\dagger (\bI + \g\trvbHe^\dagger
\trvbHe)^{-1}\trvbhr \right) \right\}^+\\
&\stackrel{\text{a.s.}}{\longrightarrow} \left\{
\log\mpf(\g,\beta) \right\}^+
\quad\text{as $\Nt\rightarrow\infty$},
\end{align*}
where as above the last line comes from applying
\eqref{eq:QuadraticForm} and \eqref{eq:etaTrans}.

\subsection{Proof of Theorem~\ref{thm:high-SNR-scaling}}
\label{sec:high-SNR-scaling}

When $\beta<1$ (i.e., $\Ne<\Nt$), we have $\bHe^\perp
\bhr\ne\mathbf{0}$ almost surely, so \eqref{eq:high-SNR-b} holds,
i.e.,
\begin{equation} 
\lim_{P\rightarrow\infty} C(P) = \infty
\end{equation}
as \eqref{eq:MISOMEscaling-C} reflects.

When $\beta\ge1$ (i.e., $\Ne>\Nt$) $\rvbHe^\dagger \rvbHe$ is
nonsingular almost surely,
\eqref{eq:high-SNR-a} holds, i.e.,
\begin{equation*}
\lim_{P\rightarrow\infty} C(P) =
\left\{\log\lambda(\rvbhr\rvbhr^\dagger,\rvbHe^\dagger
\rvbHe)\right\}^+.
\end{equation*}
Taking the limit $\Ne,\Nt\rightarrow \infty$ with $\Ne/\Nt=\beta$
fixed, and using \eqref{eq:asympF}, we obtain
\begin{equation*}
\lim_{\Nt\rightarrow\infty}\lim_{P\rightarrow\infty}
C(P) = \{-\log(\beta-1)\}^+
\end{equation*}
as \eqref{eq:MISOMEscaling-C} asserts.

Furthermore, via \eqref{eq:RMBloss} we have that
\begin{equation*} 
\lim_{P\rightarrow\infty} R_\mrm{MB}(P) =
\left\{\log\lambda(\rvbhr\rvbhr^\dagger,\rvbHe^\dagger\rvbHe)\right\}^+
= \lim_{P\rightarrow\infty} C(P),
\end{equation*}
whence \eqref{eq:MISOMEscaling-R}.

\section{Fading Channel Analysis}
\label{sec:FadingChannels} 

\iffalse

We first provide the definition of achievable rate for the fading
channel.\footnote{The realization of $\rvbhr(t)$ and $\rvbHe(t)$ is
denoted by $\bhr(t)$ and $\bHe(t)$ respectively.}

A $(n,2^{nR})$ code consists of a message set $\cW_n
=\{1,2,\ldots,2^{nR}\}$, a sequence of encoding functions $\mu_t:
\cW_n \times \compls^t \rightarrow \compls^{\Nt}$, such that at time
$t$, $\bx(t) = \mu_t(\rvw; \rvbhr^t)$ and a decoding function
$\nu(\cdot)$ that produces $\hat{\rvw} = \nu(\rvy_\mrm{r}^n,
\rvbhr^n)$.  A rate $R$ is achievable if there exists a sequence of
codes and a sequence $\eps_n$ such that $\eps_n\rightarrow 0$ as
$n\rightarrow\infty$ and (a) $\Pr(\rvw \neq \hat{\rvw}) \le \eps_n$,
(b) $\frac{1}{n}I(\rvw;\rvby_\mrm{e}^n,\rvbhr^n,\rvbHe^n) \le \eps_n$
and (c) $E\left[\sum_{t=1}^n \|\rvbx(t)\|^2\right]\le nP$.

We can convert the fading MISOME channel, which is a channel with
state $\rvbhr$ know to all terminals but state $\rvbHe$ known only to
the eavesdropper, into a memoryless channel by viewing
$(\rvby_\mrm{e},\rvbHe)$ as the eavesdropper's observation,
corresponding to expressing the channel law in the form
\begin{equation} 
p_{\rvby_\mrm{e},\rvbHe,\rvy_\mrm{r}|\rvbx,\rvbhr} 
= p_{\rvbHe} \,
p_{\rvby_\mrm{e}|\rvbHe,\rvbx}\,p_{\rvy_\mrm{r}|\rvbx,\rvbhr}.
\end{equation}
\fi

We prove the lower and upper bounds of Theorem~\ref{thm:RFF}
separately. 

\subsection{Proof of~\protect\eqref{eq:RFF-lb}}

By viewing the fading channel as a set of parallel channels indexed by the channel gain $\bhr$ 
of the intended receiver\footnote{Since the fading coefficients are continuous valued, one has to discretize these coefficients before mapping to parallel channels. By choosing appropriately fine quantization levels one can approach the rate as closely as possible. See e.g.,~\cite{khistiTchamWornell:07} for a discussion.} and the eavesdropper's observation as
$(\rvby_\mrm{e},\rvbHe)$, the rate
\begin{equation}
R = I(\rvu;\rvy_\mrm{r}\mid \rvbhr) - I(\rvu;\rvby_\mrm{e},\rvbHe\mid\rvbhr).
\label{eq:RANAcihievSI}
\end{equation}
is achievable for any choice of $p_{\rvu|\rvbhr}$ and $p_{\rvbx|\rvu,\rvbhr}$ 
that satisfies the power constraint $E[\rho(\rvbhr)]\le P$. We choose 
distributions corresponding to an adaptive version of masked beamforming,
i.e., [\cf\ \eqref{eq:MB-dists}]
\begin{equation} 
p_{\rvu |\rvbhr} = \CN(0,\tilde{\rho}(\rvbhr)),\quad p_{\rvbx|\rvu,\rvbhr} =
\CN\left(\rvu\trvbhr,\tilde{\rho}(\rvbhr)(\bI-\trvbhr\trvbhr^\dagger)\right),
\label{eq:MB-dists-FF}
\end{equation}
where we have chosen the convenient normalizations [\cf\
  \eqref{eq:tP-def} and \eqref{eq:tbhr-def}]
\begin{equation} 
\tilde{\rho}(\rvbhr) \defeq \frac{\rho(\rvbhr)}{\Nt}
\label{eq:trho-def}
\end{equation}
and 
\begin{equation} 
\trvbhr = \frac{\rvbhr}{\|\rvbhr\|}.
\label{eq:trvbhr}
\end{equation}

Evaluating \eqref{eq:RANAcihievSI} with the distributions
\eqref{eq:MB-dists-FF} yields \eqref{eq:RFF-lb} with \eqref{eq:RFF-}:
\begin{align}
I(\rvu;&\rvy_\mrm{r}\mid\rvbhr)- I(\rvu;\rvby_\mrm{e},\rvbHe\mid\rvbhr)\\
&= E[\log(1+\tilde{\rho}(\rvbhr)\|\rvbhr\|^2)]\notag\\
&\qquad {}+ E[\log(\tbhr^\dagger(\bI +
  \tilde{\rho}(\rvbhr)\rvbHe^\dagger \rvbHe)^{-1}\tbhr)]
\label{eq:RANFF2}\\ 
&= E\left[ \log\left(1+\frac{1}{\tilde{\rho}(\rvbhr)\|\rvbhr\|^2}
  \right)\right]   \notag\\
& \qquad {} + E\left[ \log\left(\tilde{\rho}(\rvbhr)\rvbhr^\dagger(\bI +
\tilde{\rho}(\rvbhr)\rvbHe^\dagger\rvbHe)^{-1}\rvbhr\right)\right],
\label{eq:RANFF3}
\end{align}
where the steps leading to~\eqref{eq:RANFF2} are analogous to those used in
Section~\ref{sec:RMB-proof} for the nonfading case and hence have been omitted.

\subsection{Proof of~\protect\eqref{eq:RFF-ub}}

Suppose that there is a sequence of $(2^{nR},n)$ codes such that for a
sequence $\eps_n$ (with $\eps_n\rightarrow 0$ as
$n\rightarrow\infty$),
\begin{equation} 
\begin{aligned}
&\frac{1}{n}H(\rvw)-\frac{1}{n}H(\rvw|\rvby_\mrm{e}^n, \rvbHe^n, \rvbhr^n) \le
 \eps_n, \\
&\Pr(\hat{\rvw}\neq \rvw) \le\eps_n.
\end{aligned}
\label{eq:codeCondn}
\end{equation}

\subsubsection{An auxiliary channel}

We now introduce another channel for which the noise vaiables $\rz_\mrm{r}(t)$ and $\rvbz_\mrm{e}(t)$ are correlated, but the conditions in~\eqref{eq:codeCondn} still hold. Hence any rate achievable on the original channel is also achievable on this new channel. In what follows, we will upper bound the rate achievable for this new channel instead of the original channel.

We begin by introducing some notation. Let,
\begin{equation}
\rho_t(\bhr^t) \defeq E\bigl[ \|\rvbx(t)\|^2 \bigm|
  \rvbhr^t=\bhr^t \bigr] 
\label{eq:powerDef}
\end{equation}
denote the transmitted power at time $t$, when the channel realization
of the intended receiver from time 1 to $t$ is
$\bhr^t$. Note that $\rho_t(\cdot)$ satisfies the long term average power constraint i.e., 
\begin{equation} 
E_{\rvbhr^n}\left[\frac{1}{n} \sum_{t=1}^n \rho_t(\rvbhr^t)
\right]\le P.\label{eq:powerCons}
\end{equation}

Next, let, $p_{\rvbhr}$ and $p_{\rvbHe}$ denote
the density functions of $\rvbhr$ and
$\rvbHe$, respectively, and  let $p_{\rvz_\mrm{r}}$ and $p_{\rvbz_\mrm{e}}$ denote the density function of the noise random variables in our channel model \eqref{eq:ChModelFF}. 
\iffalse
and let
\begin{equation*} 
p_{\rvbhr^t}(\bhr^t) = \prod_{i=1}^t
p_{\rvbhr}(\bhr(i)),
\end{equation*}
denote the density of $\rvbhr^t$ .
\fi
Observe that the constraints in \eqref{eq:codeCondn} (and hence the capacity) depend only on the
distributions $p_{\rvbz_\mrm{e}^n,\rvbhr^n, \rvbHe^n}(\bz_e^n,\bhr^n,
\bHe^n)$ and $p_{\rz_\mrm{r}^n,\rvbhr^n}(z_r^n,\bhr^n)$. Furthermore since the channel model~\eqref{eq:ChModelFF} is memoryless and $(\bhr,\bHe)$ are i.i.d.\ and mutually independent, we have
\begin{multline}
p_{\rvbz_\mrm{e}^n,\rvbhr^n, \rvbHe^n}(\bz_e^n,\bhr^n, \bHe^n) = \\
\prod_{t=1}^n p_{\rvbz_\mrm{e}}(\bz_e(t))p_{\rvbhr}(\bhr(t))p_{\rvbHe}(\bHe(t)),\label{eq:Pt1}
\end{multline}
\begin{equation}
p_{\rz_\mrm{r}^n,\rvbhr^n}(z_r^n,\bhr^n) =\prod_{t=1}^n
p_{\rvz_\mrm{r}}(z_r(t))p_{\rvbhr}(\bhr(t)).\label{eq:Pt2}
\end{equation}

Let $\cP_t$ denote the set of conditional-joint
distributions $p_{\rz_\mrm{r}(t),\rvbz_\mrm{e}(t)|\rvbhr^n,
\rvbHe^n}$ with fixed conditional-marginals, i.e., 
\begin{multline}
\cP_t = \bigl\{ p_{\rz_\mrm{r}(t),\rvbz_\mrm{e}(t)|\rvbhr^n,
  \rvbHe^n}(z_r,\bz_e \mid \bhr^n,\bHe^n ) \bigm| \\
p_{\rz_\mrm{r}(t) | \rvbhr^n,
    \rvbHe^n}(z_r \mid \bhr^n,\bHe^n)=p_{\rvz_\mrm{r}}(z_r), \\
p_{\rvbz_\mrm{e}(t)\mid  \rvbhr^n,
\rvbHe^n}(\bz_e \mid  \bhr^n,\bHe^n)=p_{\rvbz_\mrm{e}}(\bz_e)\bigr\}.
\label{eq:cPtdefn}
\end{multline}

Suppose that for each $t=1,2,\ldots, n$ we select a distribution $p_{\rz_\mrm{r}(t),\rvbz_\mrm{e}(t)|\rvbhr^n,
  \rvbHe^n} \in \cP_t$ and consider a channel with distribution
\begin{multline}
\label{eq:channelFFnew}
p_{\rz_\mrm{r}^n,\rvbz_\mrm{e}^n, \rvbhr^n,\rvbHe^n}(z_r^n,\bz_e^n,\bhr^n,\bHe^n)=\\
 \!\prod_{t=1}^n \!p_{\!\rz_\mrm{r}(t),\!\rvbz_\mrm{e}(t)|\rvbhr^n,\!
  \rvbHe^n}(\!z_r(t),\!\bz_e(t)|\bhr^n,\!\bHe^n \!)\!p_{\rvbhr}(\!\bhr(t)\!)p_{\rvbHe}(\!\bHe(t)\!).
\end{multline} This new  channel distribution has noise variables $(\rz_\mrm{r}(t),\rvbz_\mrm{e}(t))$ correlated, where the correlation is possibly time-dependent, but from~\eqref{eq:cPtdefn} and~\eqref{eq:channelFFnew}, note that $\rz_\mrm{r}^n$ and $\rvbz_\mrm{e}^n$ are marginally Gaussian and i.i.d.,  and satisfy~\eqref{eq:Pt1} and~\eqref{eq:Pt2}.  Hence the conditions in~\eqref{eq:codeCondn} are satisfied for this channel and the rate $R$ is achievable.

In the sequel we select $p_{\rz_\mrm{r}(t),\rvbz_\mrm{e}(t)|\rvbhr^n,
  \rvbHe^n}(z_r,\bz_e\mid \bhr^n,\bHe^n) $ to  be the worst case noise distribution for the Gaussian channel with gains $\bhr(t)$, and, $\bHe(t)$, 
and power of $\rho_t(\bhr^t)$ in Theorem~\ref{thm:MISOMEcap} i.e., if $\bps_t$ is the eigenvector corresponding to the largest generalized eigenvalue $\lamax(\bI+\rho_t(\bhr^t)\bhr(t)\bhr(t)^\dagger,\bI + \rho_t(\bhr^t)\bHe^\dagger(t)\bHe(t))$,
\begin{align}
&p_{\rz_\mrm{r}(t),\rvbz_\mrm{e}(t)|\rvbhr^n,
  \rvbHe^n}= \CN\left(0,\begin{bmatrix}1 & \bph_t^\dagger \\ \bph_t & \bI\end{bmatrix}\right),\label{eq:FFnoise}\text{ where }\\
&\bph_t = \begin{cases}\frac{1}{\bhr^\dagger(t)\bps_t}(\bHe(t)\bps_t),& \lambda_\mrm{max}\ge 1 ,\\
%\bHe(t)(\bHe(t)^\dagger\bHe(t))^{-1}\bhr(t), & %\lambda_\mrm{max}<1,\mrm{rank}(\bHe(t))=n_e,\\
\bGe(t)(\bGe^\dagger(t)\bGe(t))^{-1}\bgr(t),& \lambda_\mrm{max}<1,
\end{cases}\notag
\end{align}and where $\bGe(t)$ and $\bgr(t)$ are related to $\bHe(t)$ and $\bhr(t)$ as in~\eqref{eq:gG-def}. Our choice of $p_{\rz_\mrm{r}(t),\rvbz_\mrm{e}(t)|\rvbhr^n,
  \rvbHe^n}$ is such that $(\rz_\mrm{r}(t),\rvbz_\mrm{e}(t))$ only depend on the $(\rvbHe(t),\rvbhr(t),\rho_t(\rvbhr^t))$ i.e., 
\begin{equation}(\rvbHe^n,\rvbhr^n)\rightarrow (\rho(\rvbhr^t),\rvbhr(t),\rvbHe(t))\rightarrow (\rz_\mrm{r}(t),\rvbz_\mrm{e}(t))\label{eq:noiseMC}\end{equation}
forms a Markov chain.

\subsubsection{Upper bound on the auxiliary channel}
We now upper bound the secrecy rate for the channel~\eqref{eq:channelFFnew}. Note that this also upper bounds the rate on the original channel.

From Fano's inequality, that there exists a sequence $\eps'_n$
such that $\eps'_n\rightarrow 0$ as $n\rightarrow\infty$, and,
\begin{equation*} 
\frac{1}{n}H(\rvw|\rvy_\mrm{r}^n,\rvbhr^n)\le \eps'_n.
\end{equation*}\begin{align}
nR &= H(\rvw) = I(\rvw;\rvy_\mrm{r}^n \mid\rvbhr^n) + n\eps'_n\notag\\
&= I(\rvw;\rvy_\mrm{r}^n\mid\rvbhr^n) - I(\rvw;\rvby_\mrm{e}^n,\rvbHe^n\mid\rvbhr^n) +
n(\eps_n+\eps'_n)\label{eq:secrecyCondn}\\
&\le I(\rvw;\rvy_\mrm{r}^n\mid\rvbhr^n, \rvbHe^n,\rvby_\mrm{e}^n)+n(\eps_n+\eps'_n)\notag\\
&\le I(\rvbx^n;\rvy_\mrm{r}^n\mid\rvbhr^n, \rvbHe^n,\rvby_\mrm{e}^n) +n(\eps_n+\eps'_n)\label{eq:MarkovCndn1}\\
&\le\sum_{t=1}^n I(\rvbx(t);\rvy_\mrm{r}(t) \mid
\rvbHe^n,\rvbhr^n,\rvby_\mrm{e}(t))+n(\eps_n+\eps'_n)
\label{eq:memoryLess},
\end{align}
where ~\eqref{eq:secrecyCondn} follows from the secrecy condition (c.f.~\eqref{eq:codeCondn}), and \eqref{eq:MarkovCndn1} follows from the Markov relation
$\rvw\leftrightarrow (\rvbx^n, \rvby_\mrm{e}^n, \rvbhr^n,
\rvbHe^n)\leftrightarrow \rvy_\mrm{r}^n$, and \eqref{eq:memoryLess}
holds because for the channel~\eqref{eq:channelFFnew} we have
\begin{equation*}
h(\rvy_\mrm{r}^n |\rvby_\mrm{e}^n,
\rvbHe^n,\rvbhr^n, \rvbx^n ) = \sum_{t=1}^n
h(\rvy_\mrm{r}(t)|\rvby_\mrm{e}(t),\rvbhr^n,\rvbHe^n,\rvbx(t)).
\end{equation*}

We next upper bound the term $I(\rvbx(t);\rvy_\mrm{r}(t)\mid \rvby_\mrm{e}(t), \rvbHe^n, \rvbhr^n) $ in~\eqref{eq:memoryLess} for each $t=1,2,\ldots, n$.\begin{align}
&I(\rvbx(t);\rvy_\mrm{r}(t)\mid \rvby_\mrm{e}(t), \rvbHe^n, \rvbhr^n)  \notag\\
&\le I(\rvbx(t);\rvy_\mrm{r}(t)\mid \rvby_\mrm{e}(t), \rvbHe(t),\rvbhr(t),\rho_t(\rvbhr^t)) \label{eq:Hrhe} \\
&\le E[\{\log\!\lambda_\mrm{max}(\bI+\rho_t(\rvbhr^t)\rvbhr(t)\rvbhr^\dagger(t), \notag\\&\qquad\qquad\qquad  \bI + \rho_t(\rvbhr^t)\rvbHe^\dagger(t)\rvbHe(t))\}^+],\label{eq:capUB}
\end{align}
where~\eqref{eq:Hrhe} follows from the fact that (c.f.~\eqref{eq:noiseMC}),  $$(\rvbHe^n,\rvbhr^n)\rightarrow (\rvbx(t),\rho_t(\rvbhr^t),\rvbhr(t),\bHe(t))\rightarrow (\rvy_\mrm{r}(t),\rvby_\mrm{e}(t))$$forms a Markov chain
and~\eqref{eq:capUB} follows since our choice of the noise distribution in~\eqref{eq:FFnoise} is the worst case noise in~\eqref{eq:ub} for the Gaussian channel with gains $\bhr(t)$, $\bHe(t)$  and power $\rho_t(\bhr^t)$, hence the derivation in Theorem~\ref{thm:MISOMEcap} applies.

Substituting~\eqref{eq:capUB} into~\eqref{eq:memoryLess} we have,
\begin{align}
&nR - n(\eps_n + \eps'_n) \notag\\
&=\sum_{t=1}^n E_{\rvbHe(t),\rvbhr^t}\bigl[ \bigl\{ \log\lamax(\bI +
\rho_t(\rvbhr^t)\rvbhr(t)\rvbhr^\dagger(t), \notag\\
&\qquad\qquad\qquad\qquad \bI + \rho_t(\rvbhr^t)
\rvbHe^\dagger(t)\rvbHe(t)) \bigr\}^+ \bigr]\\
&\le \sum_{t=1}^n E_{\rvbHe(t),\rvbhr(t)}\bigl[ \bigl\{ \log\lamax(\bI
+ E_{\rvbhr^{t-1}}[\rho_t(\rvbhr^t)]\rvbhr(t)\rvbhr^\dagger(t),
\notag\\
&\qquad\qquad\qquad\qquad\bI +
E_{\rvbhr^{t-1}}[\rho_t(\rvbhr^t)]
\rvbHe^\dagger(t)\rvbHe(t)) \bigr\}^+ \bigr]\label{eq:concavity1}\\
&= \sum_{t=1}^n E_{\rvbHe(t),\rvbhr(t)} \bigl[ \bigl\{ \log\lamax(\bI +
\hat{\rho}_t(\rvbhr(t))\rvbhr(t)\rvbhr^\dagger(t), \notag\\
&\qquad\qquad\qquad\qquad\bI +
\hat{\rho}_t(\rvbhr(t))
\rvbHe^\dagger(t)\rvbHe(t)) \bigr\}^+ \bigr] \label{eq:defhatP}\\
&= \sum_{t=1}^n E_{\rvbHe,\rvbhr}[\{\log\lamax(\bI +
\hat{\rho}_t(\rvbhr)\rvbhr\rvbhr^\dagger, \bI + \hat{\rho}_t(\rvbhr)
\rvbHe^\dagger \rvbHe)\}^+]\label{eq:iid}\\
&\le n E_{\rvbHe,\rvbhr} \bigl[ \bigl\{ \log\lamax(\bI + \sum_{t=1}^n
\frac{1}{n}\hat{\rho}_t(\rvbhr)\rvbhr\rvbhr^\dagger, \notag\\
&\qquad\qquad\qquad\qquad\bI +
\sum_{t=1}^n\frac{1}{n}\hat{\rho}_t(\rvbhr)
\rvbHe^\dagger \rvbHe) \bigr\}^+ \bigr]\label{eq:concavity2}\\
&= n E_{\rvbHe,\rvbhr}[\{\log\lamax(\bI +
\rho(\rvbhr)\rvbhr\rvbhr^\dagger, \bI + \rho(\rvbhr) \rvbHe^\dagger
\rvbHe)\}^+]\label{eq:defP}
\end{align}
where \eqref{eq:concavity1} and \eqref{eq:concavity2} follow from
Jensen's inequality since $C(P)=\{\log\lamax(\bI +
P\rvbhr\rvbhr^\dagger, \bI + P\rvbHe^\dagger
\rvbHe)\}^+$ is a capacity and therefore concave in
$P$, \eqref{eq:defhatP} follows by defining
\begin{equation} 
\hat{\rho}_t(\rvbhr)=E_{\rvbhr^{t-1}}[\rho_t(\rvbhr^t)],\label{eq:hatRhoEq}
\end{equation}
\eqref{eq:iid} follows from the fact that the distribution of both
$\rvbhr$ and $\rvbHe$ does not depend on $t$,
and \eqref{eq:defP} follows by defining $\rho(\rvbhr) =
\frac{1}{n}\sum_{t=1}^n \hat{\rho}_t(\rvbhr)$.

To complete the proof, note that
\begin{align}
E_{\rvbhr}[\rho(\rvbhr)] \notag&=
\frac{1}{n}\sum_{t=1}^n E_{\rvbhr}[\hat{\rho}_t(\rvbhr)]\notag\\
&=\frac{1}{n}\sum_{t=1}^n E_{\rvbhr^t}[{\rho_t}(\rvbhr^t)]\label{eq:iidChannelsExp}\\
&=\frac{1}{n}\sum_{t=1}^n E_{\rvbhr^n}[{\rho_t}(\rvbhr^t)]\le P,\label{eq:powerCons2}
\end{align}
where~\eqref{eq:iidChannelsExp} follows from~\eqref{eq:hatRhoEq} and the fact that the channel gains are i.i.d., and~\eqref{eq:powerCons2} follows from~\eqref{eq:powerCons}.

\subsection{Proof of Proposition~\ref{prop:AsympRANFF}}
\label{sec:scaling-FF}

The proof is immediate from Theorems~\ref{thm:scaling},~\ref{thm:high-SNR-scaling} and~\ref{thm:RFF}.  

For the lower bound, we only consider the case when $\log\mpf(P,\beta)>0$, since otherwise the rate is zero.
We select
$\rho(\bhr) = P$ to be fixed for each $\bhr$.  Then we have from
Theorem~\ref{thm:scaling} that
\begin{equation*} 
R_{\mrm{FF},-}(\rvbhr,\rvbHe,P)
\stackrel{\text{a.s.}}{\longrightarrow} \log\mpf(P,\beta). 
\end{equation*}
Finally since almost-sure
convergence implies convergence in expectation,
$$\lim_{\Nt\rightarrow\infty} E[R_{\mrm{FF},-}(\rvbhr,\rvbHe,P)] = \log\mpf(P,\beta),$$
which establishes the lower bound~\eqref{eq:MarcenkoPasturFF}.
For the upper bound, since 
\begin{align*}
R_{\mrm{FF},+}(\bhr, \bHe, P) &= \left\{\!\log \!\lambda_\mrm{max}(\bI\!+\!
P\bh_r\bh_r^\dagger,\!\bI\! + P \bHe^\dagger\bHe\!)\!\right\}^+,
\end{align*}
we have from Theorem~\ref{thm:high-SNR-scaling} that
\begin{align}
\lim_{\Nt\rightarrow\infty}R_{\mrm{FF},+}(\bhr, \bHe, P) \stackrel{\mrm{a.s.}}{\le} \tilde{C}(\infty,\beta),
\label{eq:RFFUBas}\end{align}
and hence
\begin{align*}\lim_{\Nt\rightarrow\infty}C_{FF}(P=\g,\Nt,\Ne=\beta \Nt)
&\le\lim_{\Nt\rightarrow\infty}E[R_{FF,+}(\bhr,\bHe,\g)]\\
&\le \tilde{C}(\infty,\beta),\end{align*}
where we again use the fact that almost sure convergence implies convergence in expectation. 

\section{Concluding Remarks}
\label{sec:Conclusion}

The present work characterizes the key performance characteristics and
tradeoffs inherent in communication over the MISOME channel.  There
are many opportunities for further work.  As one example, stronger
results (i.e., tighter bounds) for the fast fading case would be quite
useful.  As another example would be extending the results to the
general MIMOME channel.  For the latter, the high SNR regime has been
characterized \cite{khistiWornellEldar:07} using generalized singular
value analysis, and the details will be reported elsewhere. 

More generally, many recent architectures for wireless systems
exploit the knowledge of the channel at the physical layer in order
to increase the system throughput and reliability.  Many of these
systems have a side benefit of providing security.  It is naturally
of interest to quantify these gains and identify potential
applications.

\section{Acknowledgement}

Yonina C.~Eldar and Ami~Wiesel provided an elegant justification that rank one covariance maximizes the
upper bound in Theorem~\ref{thm:ub}, which appears between \eqref{eq:startNumber}--\eqref{eq:entropyUB}. 

\appendices

\section{Proof of Lemma~\protect\ref{lem:ub}}
\label{app:ub}

Suppose there exists a sequence of $(2^{nR},n)$ codes such that for
every $\eps>0$, and $n$ sufficiently large we have that
\begin{align}
\Pr(\rvw \neq \hat{\rvw}) &\le \eps,\label{eq:small-err}\\
\frac{1}{n}I(\rvw;\rvby_\mrm{e}^n) &\le \eps,\label{eq:equiv-cond}\\
\frac{1}{n}\sum_{i=1}^n E[\|\rvbx(i)\|^2] &\le P. \label{eq:pow-cons}
\end{align}
We first note that \eqref{eq:small-err} implies, from Fano's
inequality,
\begin{equation}
\frac{1}{n}I(\rvw;\rvy_\mrm{r}^n) \ge R - \eps_\mrm{F}, 
\label{eq:Conv_Fano}
\end{equation}
where $\eps_\mrm{F} \rightarrow 0$ as $\eps\rightarrow 0$.
Combining \eqref{eq:equiv-cond}
and \eqref{eq:Conv_Fano}, we have for $\eps' = \eps + \eps_\mrm{F}$:
\begin{align}
nR - n\eps' &\le I(\rvw;\rvy_\mrm{r}^n) - I(\rvw;\rvby_\mrm{e}^n)\notag\\
&\le I(\rvw;\rvy_\mrm{r}^n,\rvby_\mrm{e}^n)-I(\rvw;\rvby_\mrm{e}^n)
\label{eq:R-a} \\
&= I(\rvw;\rvy_\mrm{r}^n|\rvby_\mrm{e}^n) \label{eq:R-b}\\
&= h(\rvy_\mrm{r}^n|\rvby_\mrm{e}^n) -
h(\rvy_\mrm{r}^n|\rvby_\mrm{e}^n,\rvw) \notag \\
&\le h(\rvy_\mrm{r}^n|\rvby_\mrm{e}^n) -
h(\rvy_\mrm{r}^n|\rvby_\mrm{e}^n,\rvw,\rvbx^n)\label{eq:R-c}\\ 
&= h(\rvy_\mrm{r}^n|\rvby_\mrm{e}^n) -
h(\rvy_\mrm{r}^n|\rvby_\mrm{e}^n,\rvbx^n) \label{eq:R-d}\\
&{=} h(\rvy_\mrm{r}^n|\rvby_\mrm{e}^n) -
\sum_{t=1}^n h(\rvy_\mrm{r}(t)|\rvby_\mrm{e}(t),\rvbx(t)) \label{eq:R-e}\\ 
&\le \sum_{t=1}^n h(\rvy_\mrm{r}(t)|\rvby_\mrm{e}(t)) -
\sum_{t=1}^n h(\rvy_\mrm{r}(t)|\rvby_\mrm{e}(t),\rvbx(t)) \notag\\ 
&= nI(\rvbx;\rvy_\mrm{r}|\rvby_\mrm{e},\rvq) \label{eq:R-g}\\
&\le nI(\rvbx;\rvy_\mrm{r}|\rvby_\mrm{e}), \label{eq:R-h}
\end{align}
where \eqref{eq:R-a} and \eqref{eq:R-b} each follow from the chain of
mutual information, \eqref{eq:R-c} follows from the fact that
conditioning cannot increase differential entropy, \eqref{eq:R-d}
follows from the Markov relation $\rvw \leftrightarrow (\rvbx^n,
\rvby_\mrm{e}^n) \leftrightarrow \rvy_\mrm{r}^n$, and \eqref{eq:R-e}
follows from the fact the channel is memoryless.  Moreover,
\eqref{eq:R-g} is obtained by defining a time-sharing random variable
$\rvq$ that takes values uniformly over the index set $\{1,2,\ldots,
n\}$ and defining $(\rvbx,\rvy_\mrm{r},\rvby_\mrm{e})$ to be the tuple
of random variables that conditioned on $\rvq=t$, have the same joint
distribution as $(\rvbx(t),\rvy_\mrm{r}(t),\rvby_\mrm{e}(t))$.  It
then follows that for our choice of $\rvbx$ and given
\eqref{eq:pow-cons}, $E[\|\rvbx\|^2] \le P$.  Finally, \eqref{eq:R-h}
follows from the fact that $I(\rvbx;\rvy_\mrm{r}|\rvby_\mrm{e})$ is
concave in $p_{\rvbx}$ (see, e.g., \cite[Appendix
I]{khistiTchamWornell:07} for a proof), so that Jensen's inequality
can be applied.

\section{Derivation of~\protect\eqref{eq:maxEntClaim}}
\label{app:maxEntClaim} 

The argument of the logarithm on left hand side of
\eqref{eq:maxEntClaim} is convex in $\bth$, so it is
straightforward to verify that the minimizing $\bth$ is
\begin{equation} 
\bth =(\bI + P \bHe \bHe^\dagger)^{-1}(P \bHe \bhr + \bph).
\label{eq:bth-opt-i}
\end{equation}

In the sequel, we exploit that by the definition of generalized
eigenvalues via \eqref{eq:GenEig},
\begin{equation} 
(\bI + P\bhr\bhr^\dagger)\bpsmax = \lamax(\bI + P \bHe^\dagger \bHe)\bpsmax,
\label{eq:lam-phi-char}
\end{equation}
or, rearranging,
\begin{equation} 
\left(\bhr\bhr^\dagger-\lamax\bHe^\dagger\bHe\right)\bpsmax =
\frac{(\lamax-1)}{P} \cdot \bpsmax.
\label{eq:lam-phi-char-alt}
\end{equation}

First we obtain a more convenient expression for $\bth$ as follows: 
\begin{align}
\bth &= (\bI + P \bHe \bHe^\dagger)^{-1}\left(P \bHe \bhr +
\frac{1}{\bhr^\dagger \bpsmax} \bHe\bpsmax\right) \label{eq:bth-opt-a}\\
&=(\bI + P \bHe \bHe^\dagger)^{-1}
\frac{\bHe(P\bhr\bhr^\dagger + \bI)\bpsmax}{\bhr^\dagger \bpsmax}
\notag\\
&=(\bI + P \bHe \bHe^\dagger)^{-1}
\frac{\lamax\bHe(P\bHe^\dagger \bHe + \bI)\bpsmax}{\bhr^\dagger
  \bpsmax}\label{eq:bth-opt-c} \\ 
&=(\bI + P \bHe \bHe^\dagger)^{-1}
\frac{\lamax\cdot(P\bHe\bHe^\dagger  + \bI)\bHe\bpsmax}{\bhr^\dagger
  \bpsmax} \label{eq:bth-opt-d} \\
&= \lamax \bph \label{eq:SimplifyTheta},
\end{align}
where \eqref{eq:bth-opt-a} follows from substituting 
\eqref{eq:bph-opt-I} into \eqref{eq:bth-opt-i}, and \eqref{eq:bth-opt-c}
follows from substituting via \eqref{eq:lam-phi-char}.

Next we have that 
\begin{align}
\bhr - \bHe^\dagger\bth  &= \bhr -
\frac{\lamax}{\bhr^\dagger\bpsmax}\bHe^\dagger \bHe\bpsmax 
\label{eq:diff-a}\\
&= \frac{(\bhr\bhr^\dagger -
\lamax\bHe^\dagger \bHe)\bpsmax}{\bhr^\dagger\bpsmax}\notag\\
&= \frac{(\lamax-1)\bpsmax}{P\bhr^\dagger\bpsmax}\label{eq:diff-b} \\
\end{align}
where \eqref{eq:diff-a} follows from substituting from
\eqref{eq:SimplifyTheta} with \eqref{eq:bph-opt-I}, and
\eqref{eq:diff-b} follows by substituting
\eqref{eq:lam-phi-char-alt}.   
Thus, 
\begin{equation}
P\|\bhr - \bHe^\dagger\bth\|^2 =
(\lamax-1)\left[\frac{(\lamax-1)}{P|\bhr^\dagger\bpsmax|^2}\right].
\label{eq:Pexp}
\end{equation}

To simplify \eqref{eq:Pexp} further, we exploit that
\begin{align}
1- \lamax \|\bph\|^2  
&= 1- \lamax \frac{\bpsmax^\dagger \bHe^\dagger \bHe
\bpsmax}{\bpsmax^\dagger \bhr \bhr^\dagger \bpsmax} \label{eq:ident-a} \\
&= \frac{\bpsmax^\dagger (\bhr\bhr^\dagger - \lamax
\bHe^\dagger
\bHe)\bpsmax}{|\bhr^\dagger\bpsmax|^2} \notag \\
&= \frac{(\lamax-1)}{P|\bhr^\dagger\bpsmax|^2}, \label{eq:ident-b}
\end{align}
where \eqref{eq:ident-a} follows by again substituting from
\eqref{eq:bph-opt-I}, and \eqref{eq:ident-b} follows by again
substituting from \eqref{eq:lam-phi-char-alt}.  In turn, replacing the
term in brackets in \eqref{eq:Pexp} according to \eqref{eq:ident-b} then
yields
\begin{equation} 
P\|\bhr - \bHe^\dagger\bth\|^2 = (\lamax-1) (1-\lamax\|\bph\|^2).
\label{eq:Pexp2} 
\end{equation}

Finally, substituting \eqref{eq:Pexp2} then \eqref{eq:SimplifyTheta}
into the left hand side of \eqref{eq:maxEntClaim} yields, following
some minor algebra, the right hand side as desired.

\bibliographystyle{IEEEtran}
\bibliography{sm}
\end{document}